\documentclass[runningheads]{cl2emult}
\usepackage{psfig}
\usepackage{psfrag}   
\usepackage{epsf}
\usepackage{makeidx}  
\usepackage{graphicx} 
\usepackage{subeqnar} 
\usepackage{multicol} 
\usepackage{citesort}
\usepackage{amssymb}
\usepackage{amsmath}
\usepackage{latexsym}
\usepackage{cropmark} 
\usepackage{lnp}      
\makeindex            

\begin{document}

\frontmatter



\mainmatter
\setcounter{page}{100}


\title*{{\small In: T. P\"oschel and S. Luding (eds.), {\em Granular Gases}, Lecture Notes in Physics Vol. 564, Springer (Berlin, 2000), p. 100}\vspace*{0.5cm}\\ Granular Gases with Impact-velocity Dependent Restitution Coefficient}
\label{BPpage}

\toctitle{Granular Gases with impact-velocity dependent restitution coefficient}

\titlerunning{Granular Gases with impact-velocity dependent restitution coefficient}

\author{Nikolai~V.~Brilliantov\inst{1}$^,$\inst{2} \and Thorsten P\"oschel\inst{2}}

\authorrunning{N.\,V.~Brilliantov \and T.~P\"oschel}
\tocauthor{N.\,V.~Brilliantov, T.~P\"oschel}

\institute{Moscow State University, Physics Department, Moscow 119899, Russia.\\ e-mail:~nbrillia@physik.hu-berlin.de
  \and Humboldt-Universit\"at, Institut f\"ur Physik,
  Invalidenstr. 110, D-10115 Berlin, Germany. e-mail:~thorsten@physik.hu-berlin.de, http://summa.physik.hu-berlin.de/$\sim$thorsten}

  \maketitle              

\begin{abstract}
   We consider collisional models for granular particles and analyze the 
   conditions under which the restitution coefficient might be a constant.
   We show that these conditions are not consistent with known collision 
   laws. From the generalization of the Hertz contact  law for 
   viscoelastic particles we obtain the coefficient of normal restitution 
   $\epsilon$ as a function of the normal component of the impact velocity
   $v_{\rm imp}$. Using $\epsilon(v_{\rm imp})$ we describe the 
   time evolution of temperature and of the velocity distribution function of 
   a granular gas in the homogeneous cooling regime, 
   where the particles collide according to the viscoelastic law. We show that for the studied
   systems the simple scaling hypothesis for the velocity distribution 
   function is violated, i.e. that its evolution is not determined only by 
   the time dependence of the thermal velocity. We observe, that the deviation 
   from the Maxwellian distribution, which we quantify by the value of the second 
   coefficient of the Sonine polynomial expansion of the velocity distribution 
   function,  does not depend on time 
   monotonously.  At first stage of the evolution it increases on the mean-collision 
   time-scale up to a maximum value and then decays to zero at the second stage,
   on the time scale corresponding to the evolution of the granular gas 
   temperature. For granular gas in the homogeneous cooling regime we also 
   evaluate the time-dependent self-diffusion coefficient of 
   granular particles. We analyze the time dependence of the mean-square displacement 
   and discuss its impact on clustering. Finally, we discuss 
   the problem of the relevant internal time for the systems of interest.  
\end{abstract}

\section{Introduction}

Granular gases, i.e. systems of inelastically colliding particles, are 
widely spread in nature. They may be exemplified by industrial dust or 
interterrestrial dust; the behavior of matter in planetary rings is also 
described in terms of the granular gas dynamics. As compared with common 
molecular gases, the steady removal of kinetic energy  in these systems 
due  to dissipative collisions causes a variety of nonequilibrium phenomena, 
which have been very intensively studied 
(e.g.~\cite{BSHPnumer,McNamara96,GZ93,NoijeErnst:97,EsipovPoeschel:97,NoijeErnstRing,GoldshteinShapiro95,EBEuro,NEBO97,BrilliantovPoeschel:1998d,HuthmannZippelius:97}). In most of these studies the coefficient of 
restitution, which characterizes the energy lost in the collisions, was assumed 
to be constant. This approximation, although providing a considerable 
simplification, and allowing to understand the main effects in granular gas dynamics, is not always justified (see also the paper by Thornton in this book~\cite{ThorntonHere}). Moreover, sometimes it 
occurs to be too crude to describe even qualitatively the  features of  
granular gases. Here we discuss the properties of granular gases 
consisting of viscoelastically colliding particles which implies an impact-velocity dependent restitution coefficient. The results are compared with results for gases consisting of particles which interact via a constant restitution coefficient and we see that the natural assumption of viscoelasticity leads to {\em qualitative} modifications of the gas properties.

The following problems will be addressed:
\begin{enumerate}
\item[$\bullet$]Why does the restitution coefficient $\epsilon$ depend on the
impact velocity $v_{\rm imp}$?

\item[$\bullet$]How does it depend on the impact velocity?

\item[$\bullet$]What are the consequences of the dependence of $\epsilon$
on $v_{\rm imp}$ on the collective behavior of particles in granular
gases? In particular how does $\epsilon=\epsilon\left(v_{\rm imp}\right)$ influence: 
\begin{itemize}
\item the evolution of temperature with time?
\item the evolution of the velocity distribution function with time? 
\item the self-diffusion in granular gases?
\end{itemize}
\end{enumerate}

In what follows we will show that the dependence of the restitution 
coefficient on the impact velocity is a very basic property of  
dissipative particle collisions, whereas the assumption of a constant restitution coefficient 
for the collision of three-dimensional spheres may lead to a physically incorrect dependence of the dissipative force on 
the compression rate of the colliding particles. From the Hertz 
collision law and the general relation between the elastic and dissipative
forces we deduce the dependence of the restitution coefficient on the 
impact velocity, which follows purely from scaling considerations. 
We also give the corresponding relation obtained from rigorous theory. 
Using the dependence $\epsilon(v_{\rm imp})$ we derive the time 
dependence of the temperature, the time-evolution 
of the velocity distribution function and describe self-diffusion in granular gases in the homogeneous cooling regime. 

\section{Dependence of the restitution coefficient on the impact velocity}

\begin{figure}[htbp]
\centerline{\psfig{figure=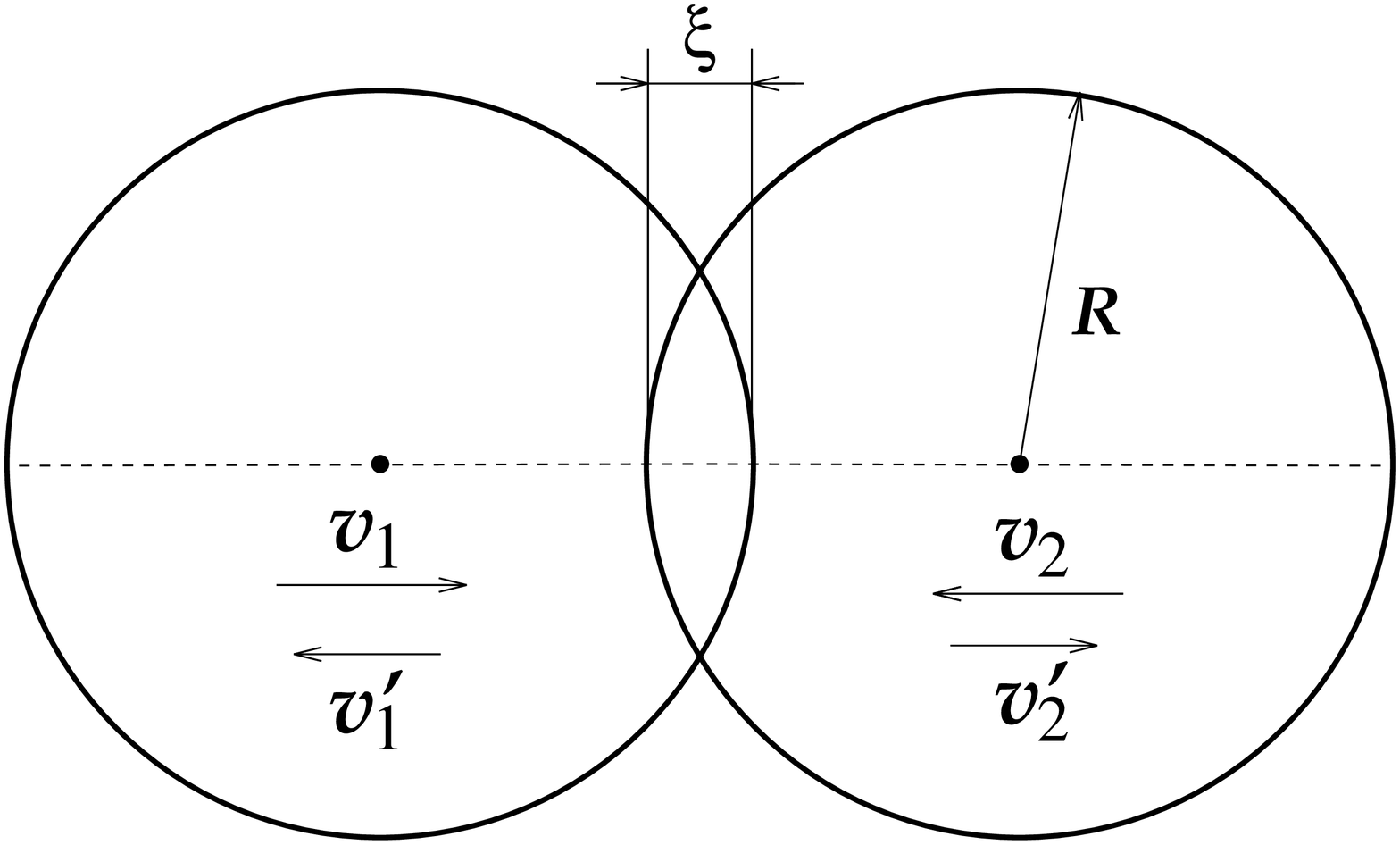,width=5.5cm}}
  \caption{Sketches of two colliding spheres. The compression $\xi$ is equal to $2R-|\vec{r}_1-\vec{r}_2|$, with $\vec{r}_{1/2}$ being the particles positions.
The compression rate is $\dot{\xi}=v_1-v_2$. For simplicity the head on 
collision of identical spheres is shown.}
  \label{fig:sketch}
\end{figure}

The collision of two particles may be characterized by the compression 
$\xi$ and by the compression rate $\dot{\xi}$, as shown on Fig.~\ref{fig:sketch}. The 
compression gives rise to the elastic force $F_{\rm el}(\xi)$, while the 
dissipative  force $F_{\rm diss}(\xi,\dot{\xi})$ appears due to the compression 
rate. 

If the compression and the compression rate are not very large, one can 
assume the dependence of the elastic and dissipative 
force on $\xi$ and $\dot{\xi}$
\begin{eqnarray}
&&F_{\rm el}(\xi) \sim \xi^{\alpha} \\
&&F_{\rm diss}(\xi,\dot{\xi}) \sim \dot{\xi}^{\beta } \xi^{\gamma }\,.
\end{eqnarray}
The dimension analysis yields the following functional 
form for the dependence of the restitution coefficient on the impact velocity
\cite{rospap}: 

\begin{equation}
\epsilon(v_{\rm imp})= 
\epsilon \left( v_{\rm imp}^{\frac{ 2 (\gamma -\alpha)}{1+\alpha}+\beta} \right)
\end{equation}
Therefore, the condition for a constant restitution coefficient imposes the 
relation between the exponents $\alpha$, $\beta$ and $\gamma$~\cite{Taguchi93,Luding98}:

\begin{equation}
2\left(\gamma -\alpha\right)+\beta\,\left ( 1+\alpha\right ) = 0\,.
\label{condition}
\end{equation}
For compressions which do not exceed the plasticity 
threshold, the particle's material behaves as a viscoelastic medium. Then 
it may be generally shown~\cite{BSHP1,BSHP2,remark2} that the  
relation 

\begin{equation}
\label{eldiss}
F_{\rm diss}= A\, \dot\xi \frac{\partial}{\partial{\xi}} F_{\rm el}(\xi) \, 
\end{equation}
between the elastic and dissipative force holds, independently 
on the shape of the bodies in contact, provided three conditions
are met~\cite{BrilPoshprep}: 

\begin{enumerate}
\label{item:condistions}
\item[(i)] The elastic components of the stress tensor 
  $\sigma_{\rm el}^{ik}$ depend linearly on
  the components of the deformation tensor $u_{ik}$~\cite{LandauLifshits}.
\item [(ii)] The dissipative components of the stress tensor 
  $\sigma_{\rm diss}^{ik}$ depend
  linearly on the components of the deformation rate tensor $\dot{u}_{ik}$~\cite{LandauLifshits}.
\item[(iii)] The conditions of quasistatic motion are provided, i.e.
  $\dot{\xi} \ll  c$, $\tau_{\rm vis} \ll \tau_c$ \cite{BSHP1,BSHP2} 
  (here $c$ is the
  speed of sound in the material of particles and $\tau_{\rm vis}$ is the 
  relaxation time of viscous processes in its bulk).

\end{enumerate}
The constant $A$ in Eq.~(\ref{eldiss}) reads~\cite{BSHP1,BSHP2}
\begin{equation}
\label{A}
A=\frac13\, \frac{(3\eta_2-\eta_1)^2}{(3\eta_2+2\eta_1)}
\left[\frac{(1-\nu^2)(1-2\nu)}{Y \nu^2}\right]\,.
\end{equation} 
where $Y$ and $\nu$ are respectively the Young modulus and the
Poisson ratio of the particle material and the viscous constants 
$\eta_1$, $\eta_2$ relate (linearly) the dissipative stress
tensor $\sigma_{\rm diss}^{ik}$ to the deformation rate tensor 
$\dot{u}_{ik}$ \cite{BSHP1,BSHP2,LandauLifshits}.

\ From Eq.~(\ref{eldiss}) follows that 
\begin{equation}
\beta=1 \qquad \gamma = \alpha -1 \,.
\label{1conditionres}
\end{equation}

Consider now a collision of three-dimensional spherical particles of 
radii $R_1$ and $R_2$. The Hertz contact contact law gives for the 
elastic force~\cite{Hertz.Brill}
\begin{equation}
\label{Hertzlaw}
F_{\rm el} =\rho \, \xi^{3/2} \,,
\qquad \rho \equiv \frac{2Y}{3(1-\nu^2)}\sqrt{R^{\rm eff}}\,,
\end{equation}
where $R^{\rm eff} \equiv R_1R_2/(R_1+R_2)$. With the set of exponents, $\alpha =3/2$, $\beta=1$ and $\gamma = 1/2$, 
which generally follows from the basic laws of the viscoelastic collision, 
the condition for the constant restitution coefficient, Eq.~(\ref{condition}), is 
obviously not satisfied. For spherical particles the restitution coefficient could be constant 
only for $\gamma = 1/4$; this, however, is not consistent with the collision laws. 
Instead one obtains the functional dependence
\begin{equation}
\epsilon=\epsilon \left( v_{\rm imp}^{1/5} \right) \,.
\label{eps15}
\end{equation}
Note that this conclusion comes from the general analysis of viscoelastic 
collisions with no other assumptions needed. Therefore, the dependence of the restitution coefficient on the impact velocity, Eq.~(\ref{eps15}), is a natural property, provided the assumption on viscoelasticity holds true which is the case in a wide range of impact velocities (see discussion on page \pageref{item:condistions}). We want to mention that the functional dependence Eq.~(\ref{eps15}) was already given in \cite{Kuwabara} using heuristic arguments.

Rigorous calculations~\cite{TomThor} yield for the dependence of the 
restitution coefficient on the impact velocity:
\begin{equation}
\label{epsC1C2}
\epsilon=1-C_1 A \kappa^{2/5}v_{\rm imp}^{1/5}
+C_2A^2 \kappa^{4/5}v_{\rm imp}^{2/5}  \mp \cdots
\end{equation}
with 
\begin{equation}
\kappa = \left( \frac32 \right)^{3/2}  
\frac{Y\sqrt{R^{\rm eff}}}{m^{\rm eff} (1-\nu^2)}
\end{equation}
where $m^{\rm eff}=m_1m_2/(m_1+m_2)$ ($m_{1/2}$ are the masses of the colliding 
particles). Numerical values for the constants $C_1$ and 
$C_2$ obtained in~\cite{TomThor} may be also written in a more convenient 
form~\cite{rospap}:

\begin{equation}
\label{C1C2}
C_1= \frac{ \Gamma(3/5)\sqrt{\pi}}{2^{1/5}5^{2/5} \Gamma(21/10)} =
1.15344, \qquad C_2=\frac{3}{5}C_1^2\, .
\end{equation}
Although the next-order coefficients of the above expansion $C_3=-0.483582$, 
$C_4=0.285279$, are now available~\cite{rospap}, we assume that 
the dissipative constant  $A$ is small enough to ignore these high-order 
terms. (For large $A$ a very accurate Pad\'e approximation for 
$\epsilon(v_{\rm imp})$ has been proposed recently~\cite{rospap}). 

\section{Time-evolution of temperature and of the velocity distribution function}

We consider a granular gas composed of $N$ identical particles 
confined in a volume $\Omega$. The particles are assumed to be smooth 
spheres, so that the collision properties are determined by the normal component of the relative 
motion only. The gas is supposed to be dilute enough so that one can 
assume binary collisions (i.e. neglect multiple collisions) and ignore the collision duration as compared with the mean free time in between successive collisions. 
We assume that the initial velocities of the 
particles (more precisely the temperature, which we define below) are not 
very large to assure viscoelastic properties of the collisions, i.e. to avoid plastic deformations and fragmentation. 
The final velocities are assumed not to be very small which allows to neglect surface forces as adhesion and others. Under these restrictions one can apply the 
viscoelastic collision model. Furthermore, we assume that dissipation is not 
large, so that the second-order expansion  (\ref{epsC1C2}) for 
$\epsilon(v_{\rm imp})$ describes the collisions accurately. We analyze the 
granular gas in the regime of homogeneous cooling, i.e. in the 
pre-clustering regime, when the gas is homogeneously distributed in space. 

The impact velocity $v_{\rm imp}$ of colliding smooth spheres, 
which determines the value of the restitution coefficient according to 
Eq.~(\ref{epsC1C2}),  is given by the normal component of the relative velocity
\begin{equation}
v_{\rm imp}=\left|\vec{v}_{12} \cdot \vec{e} \right|~~~~\mbox{with}~~~~ \vec{v}_{12}=\vec{v}_1-\vec{v}_2\,.
\end{equation}
The unit vector $\vec{e} = \vec{r}_{12}/ |\vec{r}_{12}|$ gives the 
direction of the intercenter vector $\vec{r}_{12}=\vec{r}_{1}-\vec{r}_{2}$ at 
the instant of the collision. 

The evolution of the granular gas proceeds by elementary collision events in 
which the pre-collisional velocities of colliding particles $\vec{v}_1$,  
$\vec{v}_2$, are converted into after-collisional ones, $\vec{v}_1^*$, $\vec{v}_2^*$, 
according to the rules
\begin{equation}
\label{directcoll}
\begin{split}
\vec{v}_1^*&=\vec{v}_1-
\frac12 \left[1+\epsilon(|\vec{v}_{12} \cdot \vec{e} | )\right] 
(\vec{v}_{12} \cdot \vec{e})\vec{e} \\
\vec{v}_2^*&=\vec{v}_2+ \frac12  
\left[1+\epsilon(|\vec{v}_{12} \cdot \vec{e} | )\right] 
(\vec{v}_{12} \cdot \vec{e})\vec{e} \,,
\end{split}
\end{equation}
where $\epsilon$ depends on the impact velocity 
$v_{\rm imp}=|\vec{v}_{12} \cdot \vec{e} |$. Due to the {\em direct} collision 
(\ref{directcoll}) the population in the velocity phase-space near the points 
$\vec{v}_1$,  $\vec{v}_2$ decreases, while near the points $\vec{v}_1^*$, 
$\vec{v}_2^*$ it increases. The decrease  of the population near 
$\vec{v}_1$,  $\vec{v}_2$ caused by the direct collision, is (partly) 
counterbalanced by its increase in the {\em inverse} collision, where 
the after-collisional velocities are $\vec{v}_1$,  $\vec{v}_2$ with 
the pre-collisional ones $\vec{v}_1^{**}$, $\vec{v}_2^{**}$. The 
rules for the inverse collision read
\begin{equation}
\label{inversecoll}
\begin{split}
\vec{v}_1&=\vec{v}_1^{**}-
\frac12 \left[ 1+\epsilon(|\vec{v}_{12}^{**} \cdot \vec{e} |) \right]
\left( \vec{v}_{12}^{**} \cdot \vec{e} \right)\vec{e} \\
\vec{v}_2&=\vec{v}_2^{**}+
\frac12 \left[ 1+\epsilon(|\vec{v}_{12}^{**} \cdot \vec{e} |) \right]
\left( \vec{v}_{12}^{**} \cdot \vec{e} \right)\vec{e}\,.
\end{split}
\end{equation}
Note that in contrast to the case of $\epsilon ={\rm const}$, the 
restitution coefficients in the inverse and in the direct collisions
are different.  

The Enskog-Boltzmann equation describes the evolution of the population of particles in the phase space on the mean-field level.
The evolution is characterized by the distribution 
function $f(\vec{r},\vec{v},t)$, which for the force-free case does not 
depend on $\vec{r}$ and obeys the equation~\cite{NoijeErnst:97,resibua}
\begin{multline}
\label{collint} 
\frac{\partial}{\partial t}f\left(\vec{v}_1,t\right) =
g_2(\sigma)\sigma^2 \int d \vec{v}_2 \int d\vec{e}
\Theta(-\vec{v}_{12} \cdot \vec{e}) |\vec{v}_{12} \cdot \vec{e}|\\
\times \left\{\chi f(\vec{v}_1^{**},t)f(\vec{v}_2^{**},t)-
f(\vec{v}_1,t)f(\vec{v}_2,t) \right\}  \equiv g_2(\sigma)I(f,f)
\end{multline}
where $\sigma$ is the diameter of the particles.
The contact value of the pair distribution function~\cite{CarnahanStarling}
\begin{equation}
g_2(\sigma)=(2-\eta)/2(1-\eta)^3\,,
\end{equation}
accounts on the mean-field level for the increasing frequency of collisions due to 
excluded volume effects with $\eta=\frac16\, \pi n \sigma^3$ being the 
volume fraction. 

The first term in the curled brackets in the right-hand side of 
Eq.~(\ref{collint}) refers to the ``gain'' term for the population in the 
phase-space near  the point $\vec{v}_1$, while the second one is the 
``loss'' term. The Heaviside function $\Theta(-\vec{v}_{12} \cdot \vec{e})$ discriminates approaching particles (which do collide) from separating particles (which do not collide), and
$|\vec{v}_{12} \cdot \vec{e}|$ gives the length of the collision cylinder.
Integration in Eq.~(\ref{collint}) is performed over all 
velocities $\vec{v}_2$ and interparticle vectors  $\vec{e}$ in the direct 
collision. Equation~(\ref{collint}) accounts also 
for the inverse collisions via the factor $\chi$, which appears due to 
the Jacobian of the transformation 
$\vec{v}_1^{**},\vec{v}_2^{**} \to  \vec{v}_1,\vec{v}_2$, and due to the 
difference between the lengths of the collision cylinders of  the 
direct and the inverse collision:
\begin{equation}
\chi=
\frac{{\cal D}(\vec{v}_1^{**},\vec{v}_2^{**} )}{{\cal D}(\vec{v}_1,\vec{v}_2)}
\, 
\frac{|\vec{v}_{12}^{**} \cdot \vec{e}|}{|\vec{v}_{12} \cdot \vec{e}|}\,.
\end{equation}

For constant restitution coefficient the factor $\chi$ is a 
constant
\begin{equation}
\chi = \frac{1}{\epsilon^2} = {\rm const} \, , 
\end{equation}
while for $\epsilon=\epsilon\left(v_{\rm imp}\right)$, as given in Eq.~(\ref{epsC1C2}), it reads~\cite{BPMaxw}
\begin{equation}
\label{chi}
\chi = 1 + \frac{11}{5}C_1 A \kappa^{2/5} |\vec{v}_{12} \cdot \vec{e}|^{1/5}
+\frac{66}{25}C_1^2 A^2 \kappa^{4/5} |\vec{v}_{12} \cdot \vec{e}|^{2/5} 
+\cdots
\end{equation}
\ From Eq.~(\ref{chi}) it follows that $\chi =\chi(|\vec{v}_{12} \cdot \vec{e}|)$. 
Since the average velocity in granular gases changes with time, such a
dependence of $\chi$ means, as we will show below, that $\chi$ and, therefore, the velocity distribution function itself depend explicitly on time.
The time dependence of $\chi$ changes drastically the properties of the collision integral 
and destroys the simple scaling form of the velocity distribution function, 
which holds for the case of the constant restitution coefficient 
(e.g.~\cite{EsipovPoeschel:97,NoijeErnst:97}). 

Nevertheless, some important properties of the collision integral are preserved. Namely, it may be shown that the relation 
\begin{eqnarray}
\label{deraver} 
&&\!\!\!\!\!\!\frac{d}{dt} \left< \psi(t) \right> =\int d\vec{v}_1 \psi (\vec{v}_1)
\frac{\partial}{\partial t} f(\vec{v}_1, t) = g_2(\sigma)
\int d\vec{v}_1 \psi (\vec{v}_1) I(f,f) =\\
&&\!\!\!\!\!\!\frac{g_2(\sigma)\sigma^2}{2} \int d\vec{v}_1d\vec{v}_2 \int d\vec{e}\,
\Theta(-\vec{v}_{12} \cdot \vec{e}) |\vec{v}_{12} \cdot \vec{e}|
f(\vec{v}_1, t)f(\vec{v}_2, t) \Delta \left[ \psi( \vec{v}_1)+
\psi( \vec{v}_2) \right] \nonumber 
\end{eqnarray}
holds true, where 
\begin{equation}
\left< \psi(t) \right> \equiv \int d\vec{v} \psi (\vec{v}) f(\vec{v}, t)
\end{equation}
is the average of some function $\psi (\vec{v})$, and 
\begin{equation}
\Delta \psi(\vec{v}_i) \equiv \left[\psi(\vec{v}_i^*)-\psi(\vec{v}_i) \right]
\end{equation}
denotes the change of $\psi( \vec{v}_i)$ in a direct collision. 

Now we introduce the temperature of the three-dimensional granular gas,
\begin{equation}
\label{deftemp1}
\frac{3}{2} n T(t)=\int d \vec{v} \frac{m v^2}{2} f(\vec{v},t)\,,
\end{equation}
where $n$ is the number density of granular particles 
($n=N/ \Omega$), and the characteristic velocity $v_0^2(t)$ is related to temperature via
\begin{equation}
\label{deftemp}
T(t)=\frac12 mv_0^2(t)\,.
\end{equation}
First we try the scaling ansatz
\begin{equation}
\label{veldis}
f(\vec{v}, t)=\frac{n}{v_0^3(t)} \tilde{f}(\vec{c})
\end{equation}
where 
$\vec{c} \equiv \vec{v}/v_0(t)$ and following 
\cite{NoijeErnst:97,GoldshteinShapiro95} assume that deviations from the 
Maxwellian distribution are not large, so that  $\tilde{f}(\vec{c})$ may 
be expanded into a convergent series with the leading term being the 
Maxwellian distribution $\phi(c) \equiv \pi^{-3/2} \exp(-c^2)$.  
It is convenient to use the Sonine polynomials expansion 
\cite{NoijeErnst:97,GoldshteinShapiro95}
\begin{equation}
\label{Soninexp}
\tilde{f}(\vec{c})=\phi(c) \left\{1 + \sum_{p=1}^{\infty} a_p S_p\left(c^2\right) \right\}
\, .
\end{equation}
These polynomials are orthogonal, i.e.
\begin{equation}
\label{Soninortog}
\int d \vec{c}\, \phi (c) S_p(c^2)S_{p^{\prime}}\!\left(c^2\right) 
= \delta_{pp^{\prime}}{\cal N}_p\,,
\end{equation}
where $\delta_{pp^{\prime}}$ is the Kronecker delta and  ${\cal N}_p$ is the 
normalization constant. The first few polynomials  read
\begin{equation}
\label{Soninfewfirst}
\begin{split}
S_0(x)&=1 \\
S_1(x)&=-x^2 +\frac32 \\
S_2(x)&=\frac{x^2}{2}-\frac{5x}{2}+\frac{15}{8}\,.
\end{split}
\end{equation}
Writing the Enskog-Boltzmann equation in terms of the scaling variable
$\vec{c}_1$, one observes that the factor $\chi$ may not be expressed 
only in terms of the scaling variable, but it depends also on the 
characteristic velocity $v_0(t)$, and thus depends on time. Therefore, the 
collision integral also occurs to be time-dependent. As a result, it is not 
possible to reduce 
the Enskog-Boltzmann equation to a pair of equations, 
one for the time evolution of the temperature and another for the 
time-independent scaling function, whereas for $\epsilon=$const. the Boltzmann-Enskog equation is separable, e.g.~\cite{NoijeErnst:97,GoldshteinShapiro95,EsipovPoeschel:97}. Formally adopting the approach of Refs.~\cite{NoijeErnst:97,GoldshteinShapiro95} for $\epsilon=$const., one would obtain  
time-dependent coefficients $a_p$ of the Sonine polynomials expansion. 
This means that the simple scaling ansatz (\ref{veldis}) is violated 
for the case of the impact-velocity dependent restitution coefficient. 

Thus, it seems natural to write the distribution function in the following 
general form  

\begin{equation}
\label{genscal}
f(\vec{v}, t)=\frac{n}{v_0^3(t)}\tilde{f}(\vec{c}, t)
\end{equation}
with 
\begin{equation}
\label{genSoninexp}
\tilde{f}(\vec{c})
=\phi(c) \left\{1 + \sum_{p=1}^{\infty} a_p(t) S_p(c^2) \right\}
\end{equation}
and find then equations for the {\em time-dependent} coefficients $a_p(t)$. 
Substituting (\ref{genscal}) into the Boltzmann equation (\ref{collint})
we obtain
\begin{equation}
\label{geneqveldis}
\frac{\mu_2}{3} 
\left(3 + c_1 \frac{\partial}{\partial c_1} \right) \tilde{f}(\vec{c}, t) +
B^{-1} \frac{\partial}{\partial t} \tilde{f}(\vec{c}, t) =
\tilde{I}\left( \tilde{f}, \tilde{f} \right)
\end{equation}
with
\begin{equation}
B=B(t) \equiv v_0(t) g_2(\sigma) \sigma^2 n\,.
\end{equation}
We define the 
dimensionless collision integral:
\begin{multline}
\label{dimlcolint}
\tilde{I}\left( \tilde{f}, \tilde{f} \right)=\\
\int d \vec{c}_2 \int d\vec{e}
\Theta(-\vec{c}_{12} \cdot \vec{e}) |\vec{c}_{12} \cdot \vec{e}|
\left\{\tilde{\chi} \tilde{f}(\vec{c}_1^{**},t) \tilde{f}(\vec{c}_2^{**},t)-
\tilde{f}(\vec{c}_1,t)\tilde{f}(\vec{c}_2,t) \right\} 
\end{multline}
with the reduced factor $\tilde{\chi}$
\begin{equation}
\label{chiscal}
\tilde{\chi} = 1 + 
\frac{11}{5}C_1 \delta^{\prime} |\vec{c}_{12} \cdot \vec{e}|^{1/5}
+\frac{66}{25}C_1^2 \delta^{\prime \, 2} |\vec{c}_{12} \cdot \vec{e}|^{2/5}
+\cdots
\end{equation}
which depends now on time via a quantity
\begin{equation}
\label{deltaprime}
\delta^{\, \prime} (t) \equiv A \kappa^{2/5} \left[2T(t)\right]^{1/10} 
\equiv \delta \left[2T(t)/T_0 \right]^{1/10}\,. 
\end{equation}
Here $\delta \equiv A \kappa^{2/5}[T_0]^{1/10}$, $T_0$ is the initial 
temperature, and for simplicity we assume the unit mass, $m=1$. 
We also define the moments of the dimensionless collision integral
\begin{equation}
\label{mup}
\mu_p \equiv - \int d \vec{c}_1 c_1^p 
\tilde{I}\left( \tilde{f}, \tilde{f} \right)\ , 
\end{equation}
so that the second moment describes the rate of the temperature change:
\begin{equation}
\label{dTdt}
\frac{dT}{dt} =-\frac23 BT\mu_2\,.
\end{equation} 
Equation~(\ref{dTdt}) follows from the definitions of the temperature and of the 
moment $\mu_2$. Note that these moments depend on time, in contrast to the 
case of the constant restitution coefficient, where these moments are time-independent~\cite{NoijeErnst:97}. 

Multiplying both sides of Eq.~(\ref{geneqveldis}) with $c_1^p$ and integrating 
over $d \vec{c}_1$, we obtain
\begin{equation}
\label{momeq}
\frac{\mu_2}{3} p \left< c^p \right> -B^{-1}\sum_{k=1}^{\infty} 
\dot{a}_k \nu_{kp} = \mu_p
\end{equation}
where integration by parts has been performed and we define
\begin{eqnarray}
\label{nukp}
\nu_{kp} &\equiv& \int \phi(c) c^p S_k(c^2) d\vec{c} \\
\left< c^p \right>  &\equiv& \int c^p \tilde{f}(\vec{c}, t)  d\vec{c}\,.
\label{cpdef}
\end{eqnarray}
The calculation of $\nu_{kp}$ is straightforward; the first few of these  read:
$\nu_{22}=0$, $\nu_{24}=\frac{15}{4}$. The odd moments 
$\left< c^{2n+1} \right> $ vanish, while the even ones 
$\left< c^{2n} \right> $ may be expressed in terms of $a_k$ with 
$0 \leq k  \leq n$, namely, 
$\left< c^2 \right>  =\frac32 - \frac32 a_1$. On the other hand, from the
definition of temperature and of the thermal velocity in Eqs.~(\ref{deftemp}) and (\ref{deftemp1})
follows that $\left< c^2 \right> = \frac32$ and thus, $a_1=0$. 
Similar considerations yield 
$\left< c^4 \right> = \frac{15}{4}\left( 1 + a_2 \right)$. 
The moments $\mu_p$ may be expressed in terms of coefficients 
$a_2, a_3, \cdots$ too; therefore, the system Eq.~(\ref{momeq}) is an infinite (but closed) set of equations for these 
coefficients. 

It is not possible to get a general solution of the problem. However, since
the dissipative parameter $\delta$ is supposed to be small, the deviations from the 
Maxwellian distribution are presumably small too. Thus, we assume that one can neglect all  
high-order terms with $p>2$ in the expansion (\ref{genSoninexp}). Then Eq.~(\ref{momeq}) 
is an equation for the coefficient $a_2$. For $p=2$ Eq.~(\ref{momeq}) converts into an identity,
since $\left< c^2 \right>=\frac32$, $a_1=0$, $\nu_{22}=0$ and 
$\nu_{24}=\frac{15}{4}$. For $p=4$ we obtain
\begin{equation}
\label{eqa2}
\dot{a}_2-\frac43\, B\mu_2 \left(1+a_2 \right)+\frac{4}{15}B\mu_4 =0\,.
\end{equation} 
In Eq.~(\ref{eqa2}) $B$ 
depends on time as 
\begin{equation}
B(t)=(8 \pi)^{-1/2} \tau_c(0)^{-1}[T(t)/T_0]^{1/2}\,,
\end{equation}
where $\tau_c(0)$ is related to the initial mean-collision time, 
\begin{equation}
\tau_c(0)^{-1}=4 \pi^{1/2}g_2(\sigma) \sigma^2 n T_0^{1/2}\,.
\end{equation}
The time evolution of the temperature is determined by Eq.~(\ref{dTdt}), i.e. by the 
time dependence of $\mu_2$. 

The time-dependent coefficients $\mu_p(t)$ may be expressed in terms of $a_2$ owing to 
their definition Eq.~(\ref{mup}) and the approximation $\tilde{f}= \phi (c) [ 1+a_2(t)S_2(c^2)]$. 
We finally obtain:
\begin{multline}
\label{mupa2}
\mu_p=-\frac12 \int d\vec{c}_1\int d\vec{c}_2 \int d\vec{e} 
\Theta(-\vec{c}_{12} \cdot \vec{e}) |\vec{c}_{12} \cdot \vec{e}| \phi(c_1) \phi(c_2)\\ \times 
\left\{1+a_2\left[S_2(c_1^2)+S_2(c_2^2) \right] + a_2^2\,S_2(c_1^2)S_2(c_2^2) \right\}
\Delta (c_1^p+c_2^p) 
\end{multline}
with the definition of $\Delta (c_1^p+c_2^p)$ given above. Calculations 
performed up to the second order in terms of the dissipative parameter $\delta$ 
yield~\cite{BPMaxw}:
\begin{equation}
\label{mu2A}
\mu_2= \sum_{k=0}^{2} \sum_{n=0}^2 {\cal A}_{kn} \delta^{\, \prime \,k } a_2^n
\end{equation} 
where
\begin{align}
\label{A1A6}
{\cal A}_{00}&=0; \quad &{\cal A}_{01}&=0; \quad &{\cal A}_{02}&=0 \nonumber\\
{\cal A}_{10}&=\omega_0; \quad &{\cal A}_{11}&=\frac{6}{25}\omega_0; \quad &{\cal A}_{12}&=\frac{21}{2500} \omega_0 \\
{\cal A}_{20}&=-\omega_1; \quad & {\cal A}_{21}&=-\frac{119}{400}\omega_1; \, \quad &{\cal A}_{22}&=-\frac{4641}{640000}\omega_1\nonumber
\end{align}
with 
\begin{eqnarray}
\omega_0 &\equiv& 2 \sqrt{2 \pi} 2^{1/10} \Gamma \left (\frac{21}{10} \right)C_1=6.48562 \ldots\\
\omega_1 &\equiv& \sqrt{2 \pi} 2^{1/5} \Gamma \left (\frac{16}{5} \right)C_1^2=9.28569 \ldots
\end{eqnarray}
Similarly  
\begin{equation}
\label{mu4A}
\mu_4= \sum_{k=0}^{2} \sum_{n=0}^2 {\cal B}_{kn} \delta^{\, \prime \,k } a_2^n
\end{equation} 
with

\begin{align}
{\cal B}_{00}&=0; \quad &{\cal B}_{01}&=4\sqrt{2 \pi}; \quad &{\cal B}_{02}&=\frac{1}{8}\sqrt{2 \pi} \nonumber \\
\label{B4B6}
{\cal B}_{10}&= \frac{56}{10}\omega_0; \quad  &{\cal B}_{11}&=\frac{1806}{250}\omega_0; \quad &{\cal B}_{12}&=\frac{567}{12500}\omega_0 \\
{\cal B}_{20}&= -\frac{77}{10}\omega_1;\quad  &{\cal B}_{21}&=-\frac{149054}{13750}\omega_1;\quad &{\cal B}_{22}&=-\frac{348424}{5500000}\omega_1\nonumber
\end{align}

Thus, Eqs.~(\ref{dTdt}) and (\ref{eqa2}), together with Eqs.~(\ref{mu2A}) and  
(\ref{mu4A}) form a  closed set to find the time evolution of the temperature and 
the coefficient $a_2$. We want to stress an important difference for the time evolution 
of temperature for the case of the impact-velocity dependent restitution coefficient, 
as compared to that of a constant restitution coefficient. 
In the former case it is coupled to the time evolution of the coefficient $a_2$, while  
in the latter case there is no such coupling since $a_2={\rm const}$. 
This coupling may lead in to a rather peculiar time-dependence of 
the temperature. 

Introducing  the reduced temperature $u(t) \equiv T(t)/T_0$ we recast the set of equations
(\ref{dTdt}) and (\ref{eqa2}) into the following form:
\begin{multline}
\label{genseteq1}
\dot{u}+\tau_0^{-1}u^{8/5}\left( \frac53 +\frac25 a_2+\frac{7}{500}a_2^2 \right)\\
-\tau_0^{-1}q_1 \delta \, u^{17/10}
\left( \frac53 +\frac{119}{240}a_2 +\frac{1547}{128000}a_2^2 \right) =0
\end{multline}
\begin{equation}
\label{genseteq2}
\dot{a}_2-r_0u^{1/2}\mu_2 \left(1 +a_2 \right) + \frac15 r_0u^{1/2}\mu_4=0\,,
\end{equation}
where we introduce the characteristic time  
\begin{equation}
\label{tau0}
\tau_0^{-1}=\frac{16}{5} q_0 \delta \cdot \tau_c(0)^{-1}
\end{equation} 
with
\begin{eqnarray}
q_0&=&2^{1/5}\Gamma(21/10)C_1/8=5^{-2/5}\sqrt{\pi}\Gamma(3/5)/8=0.173318\ldots \\
r_0 &\equiv& \frac{2}{3 \sqrt{2 \pi}} \tau_c(0)^{-1}\\
q_1 &\equiv& 2^{1/10} (\omega_1/\omega_0) =1.53445 \ldots
\end{eqnarray}
As shown below the characteristic time $\tau_0$ describes the time evolution of the temperature.
To obtain these equations we use the 
expressions for $\mu_2(t)$, $B(t)$, and for the coefficients ${\cal A}_{nk}$. Note that the 
characteristic time $\tau_0$ is $\delta^{-1} \gg 1$ times larger than the collision 
time $ \sim \tau_c(0)$. 

We will find the  solution to these equations as expansions in terms of the small 
dissipative parameter $\delta$ ($\delta^{\, \prime}(t) = \delta \cdot 2^{1/10}u^{1/10}(t)$):
\begin{eqnarray}
\label{expudel}
&&u=u_0+ \delta \cdot u_1 +\delta^2 \cdot u_2 +\cdots \\
\label{expadel}
&&a_2=a_{20}+\delta \cdot a_{21}+\delta^2 \cdot a_{22} +\cdots
\end{eqnarray} 
Substituting Eqs.~(\ref{mu2A},\ref{mu4A},\ref{expudel},\ref{expadel}) into 
Eqs.~(\ref{genseteq1},\ref{genseteq2}), one can solve these equations perturbatively, 
for each order of $\delta$. The solution of the order of ${\cal O}(1)$ reads for
the coefficient $a_2(t)$ \cite{BPMaxw}:
\begin{equation}
\label{a20tsmall}
a_{20}(t) \approx a_{20}(0)e^{-4t\left/\left(5 \tau_{\rm E}(0)\right)\right.}\,,
\end{equation} 
where $\tau_{\rm E} =\frac32 \tau_c$ is the Enskog relaxation time, so that 
$a_{20}(t)$ vanishes for $t \sim \tau_0$. This refers to the relaxation of an initially non-Maxwellian velocity distribution to the Maxwellian distribution. Note that the relaxation occurs within few collisions per particle, similarly to the relaxation of common molecular gases. 

We now assume that the initial distribution is Maxwellian, i.e., that $a_{20}(0)=0$ for 
$t=0$. Then the deviation from the Maxwellian distribution originates from the 
inelasticity of the interparticle collisions. For the case of  $a_{20}(0)=0$ the 
solution of the order of ${\cal O}(1)$ for the reduced temperature reads
\begin{equation}
\label{T(t)del1}
\frac{T(t)}{T_0}=u_0(t)= \left(1 + \frac{t}{\tau_0} \right)^{-5/3}\,,
\end{equation} 
which coincides with the time-dependence of the temperature obtained previously 
using scaling arguments~\cite{TomThor} (up to a constant $\tau_0$ which may 
not be determined by scaling arguments). 

The solution for $a_2(t)$ in linear approximation with respect to $\delta$ reads
\begin{equation}
\label{a21gensolLi}
a_2(t)=\delta \cdot a_{21}(t)=-\frac{12}{5}w(t)^{-1}
\left\{ {\rm Li} \left[ w(t) \right]-{\rm Li} \left[ w(0) \right] \right\}
\end{equation} 
where 
\begin{equation}
\label{w(t)}
w(t) \equiv \exp \left[ \left(q_0 \delta \right)^{-1} \left(1+t/\tau_0 \right)^{1/6} \right]\,.
\end{equation} 
and with the logarithmic Integral
\begin{equation}
{\rm Li}(x)=\int\limits_0^x\frac{1}{\ln(t)} dt\,.
\end{equation}
For $t \ll \tau_0$ the coefficient $a_2(t)$ 
(\ref{a21gensolLi}) reduces to 
\begin{equation}
\label{a21tsmallsol}
a_{2}(t)=
- \delta \cdot h \left( 1- e^{-4t\left/\left(5\tau_{\rm E}(0)\right)\right.} \right)
\end{equation} 
where 
\begin{equation}
h \equiv 2^{1/10} \left( {\cal B}_{10}-5{\cal A}_{10}\right)/16 \pi 
=(3/10)\Gamma(21/10)2^{1/5}C_1=0.415964\,.
\end{equation}
As it follows from Eq.~(\ref{a21tsmallsol}), after a transient 
time of the order of few collisions per particle, i.e. for 
$\tau_{\rm E}(0) < t \ll \tau_0$, $a_{2}(t)$ saturates at the 
``steady-state''- value $-h\, \delta=-0.415964\,\delta$, i.e. it changes only slowly on the time-scale $\sim \tau_c(0)$. On the other hand, for $t \gg \tau_0$ one obtains
\begin{equation}
\label{a21tlargesol}
a_{2}(t) \simeq
- \delta \cdot h \left( t/\tau_0  \right)^{-1/6}
\end{equation} 
so that $a_{2}(t)$ decays to zero on a 
time-scale $\sim \tau_0$, i.e. slowly in the collisional time-scale 
$\sim \tau_c(0) \ll \tau_0$. The velocity distribution thus tends asymptotically 
to the Maxwellian distribution. For this regime the first-order correction 
for the reduced temperature, $u_1(t)$, reads~\cite{BPMaxw}:
\begin{equation}
\label{u1asympsol}
u_1(t)=\left( \frac{12}{25}h+2q_1 \right)(t/\tau_0)^{-11/6}=
3.26856\,(t/\tau_0)^{-11/6}\,,
\end{equation}
where we used the above results for the constants $h$ and $q_1$. From the last 
equation one can see  how the coupling between the temperature evolution and the 
evolution of the velocity distribution influences the evolution of temperature. Indeed, if there 
were  no such coupling, there would be no coupling term in Eq.~(\ref{genseteq1}), and 
thus no contribution from $\frac{12}{25}h$ to the prefactor of $u_1(t)$ in Eq.~(\ref{u1asympsol}). This would noticeably change  the time behavior of $u_1(t)$. 
On the other hand, the leading term in the time dependence of temperature, 
$u_0(t)$, is not affected by this kind of coupling. 

\begin{figure}[htbp]
\centerline{\psfig{figure=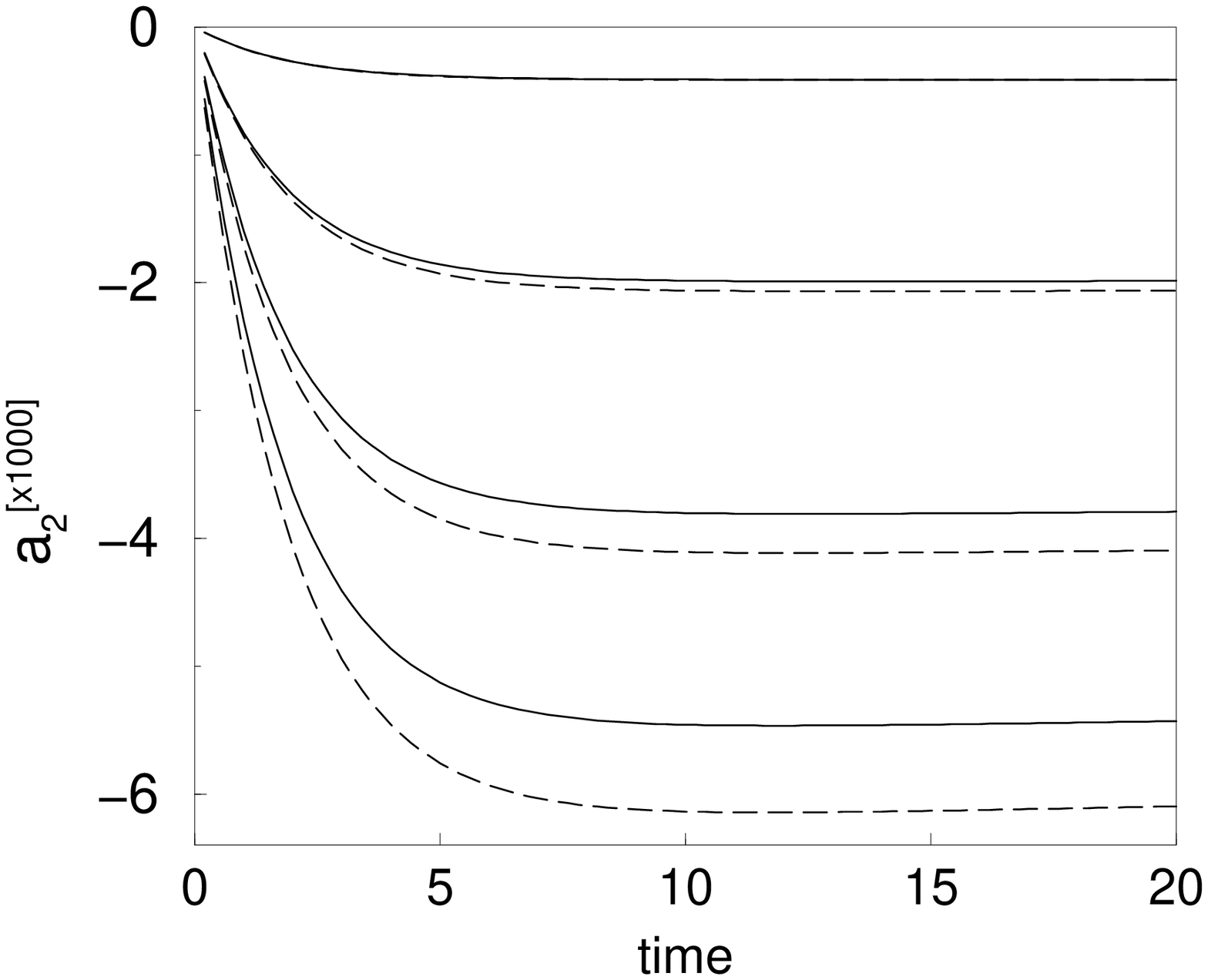,width=5.5cm}\psfig{figure=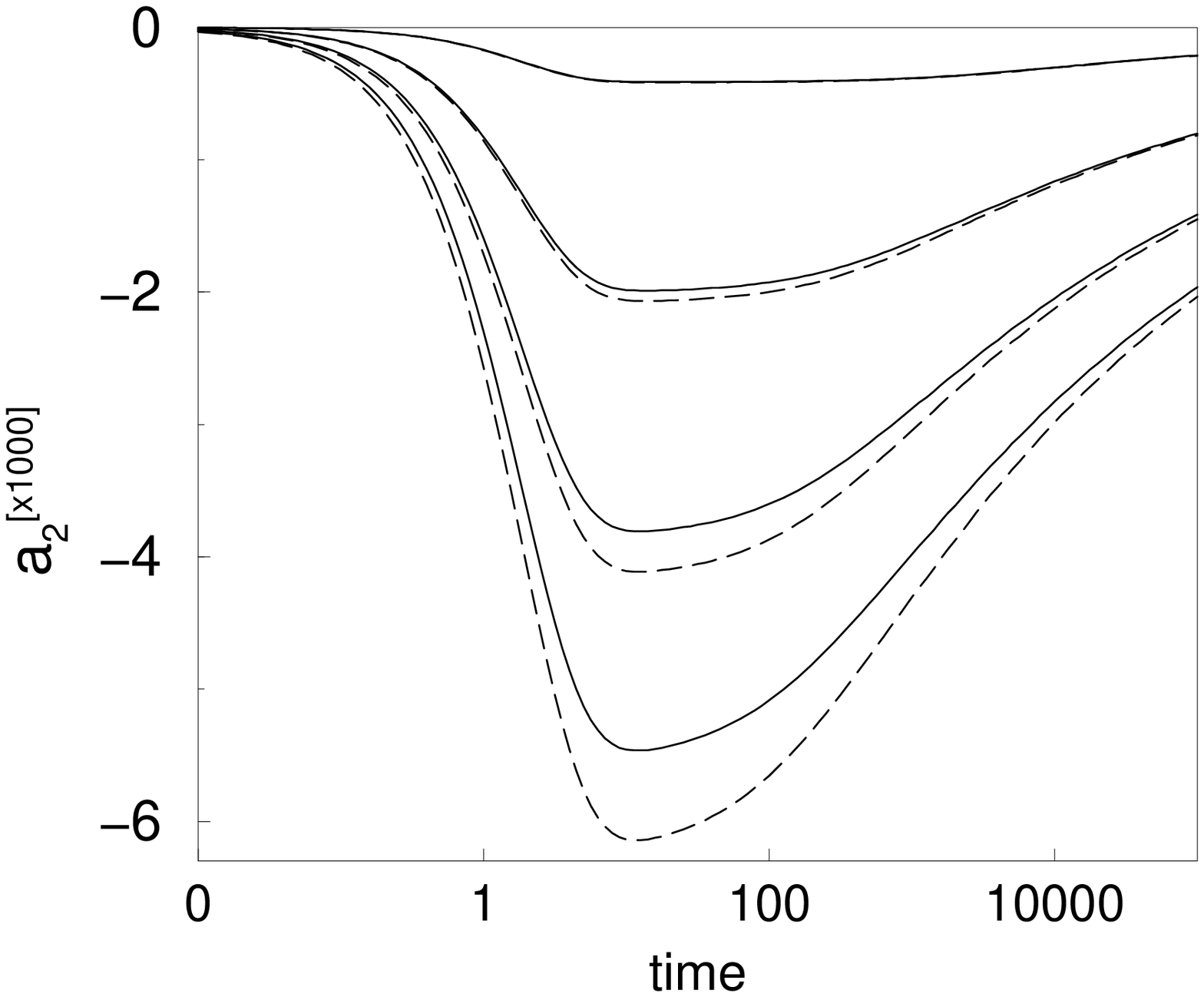,width=5.5cm}}
\centerline{\psfig{figure=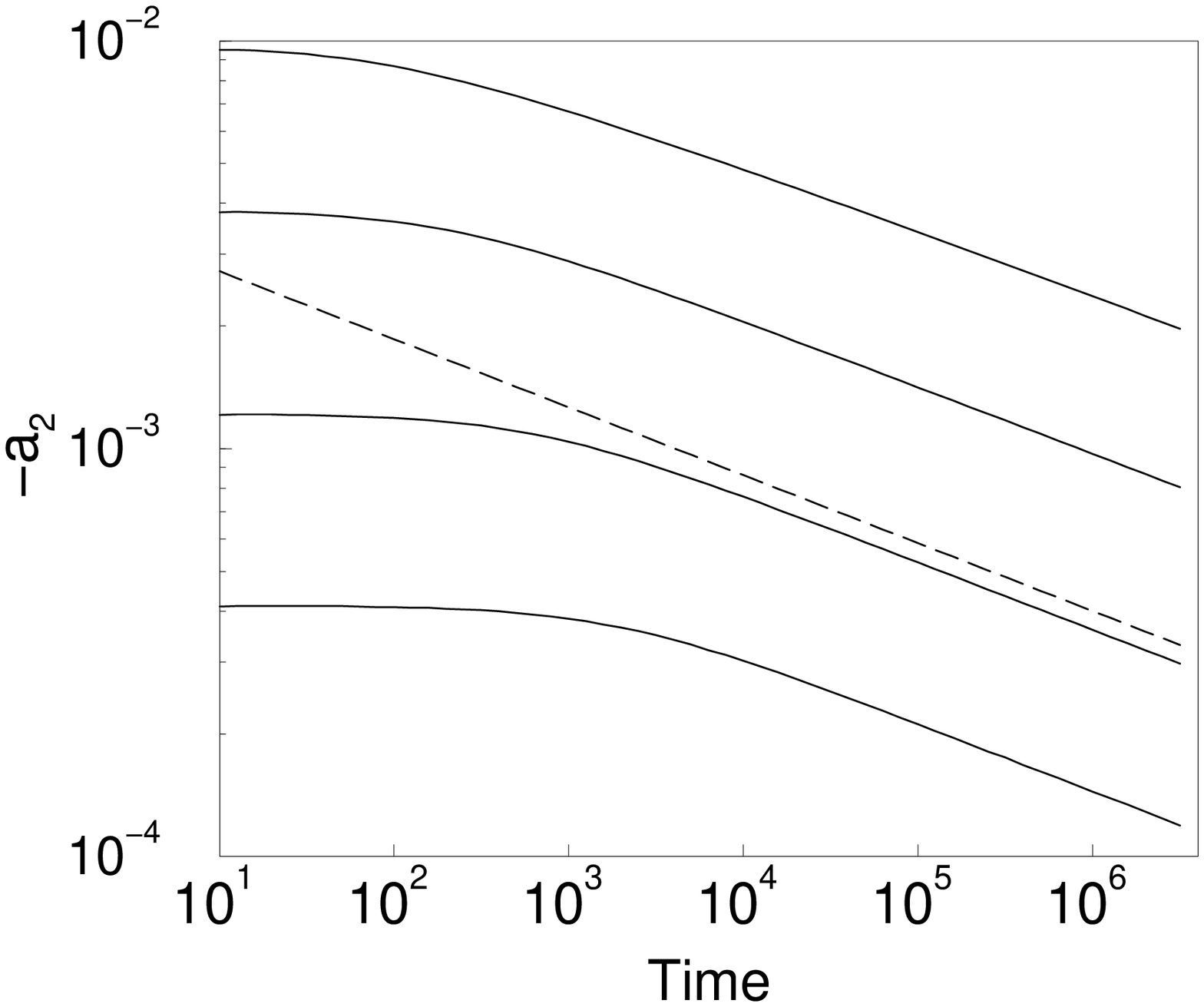,width=5.5cm}}
  \caption{Time dependence of the second coefficient of the Sonine polynomial 
expansion $a_2(t)$. Time is given in units of the mean collisional time 
$\tau_c(0)$. (Left): $a_2 \times 1000$ (solid lines) for 
$\delta =0.001, 0.005, 0.01, 0.015$ 
(top to bottom) together with the linear approximation (dashed lines); (Right): the same as (left) but for larger times; (Middle): $-a_2(t)$ over time (log-scale) for $\delta =0.03, 0.01, 0.003, 0.001$ (top to bottom) together with the power-law asymptotics $\sim t^{-1/6}$.}
  \label{fig:a2}
\end{figure}

\begin{figure}[htbp]
\centerline{\psfig{figure=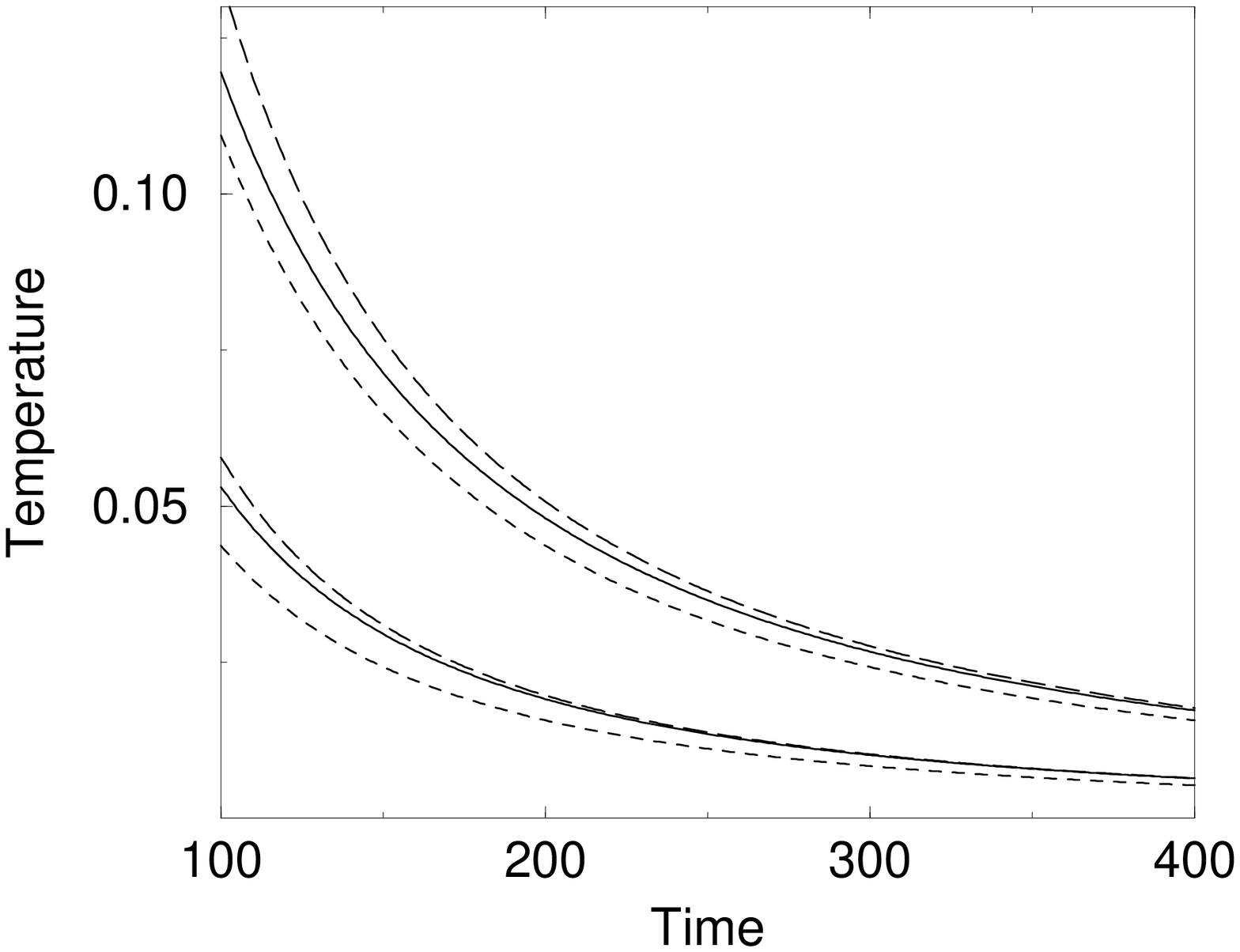,width=5.5cm}\psfig{figure=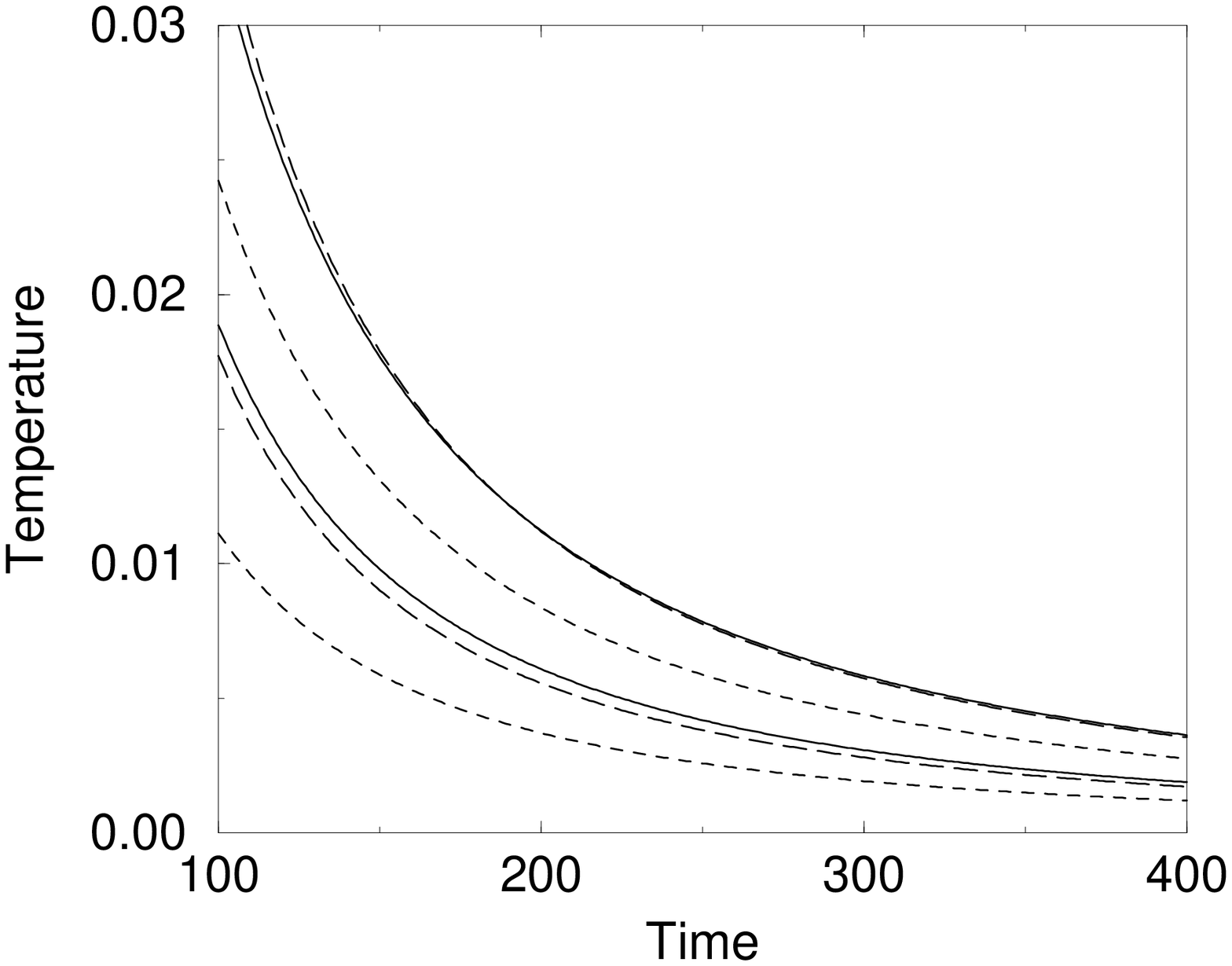,width=5.5cm}}

\centerline{\psfig{figure=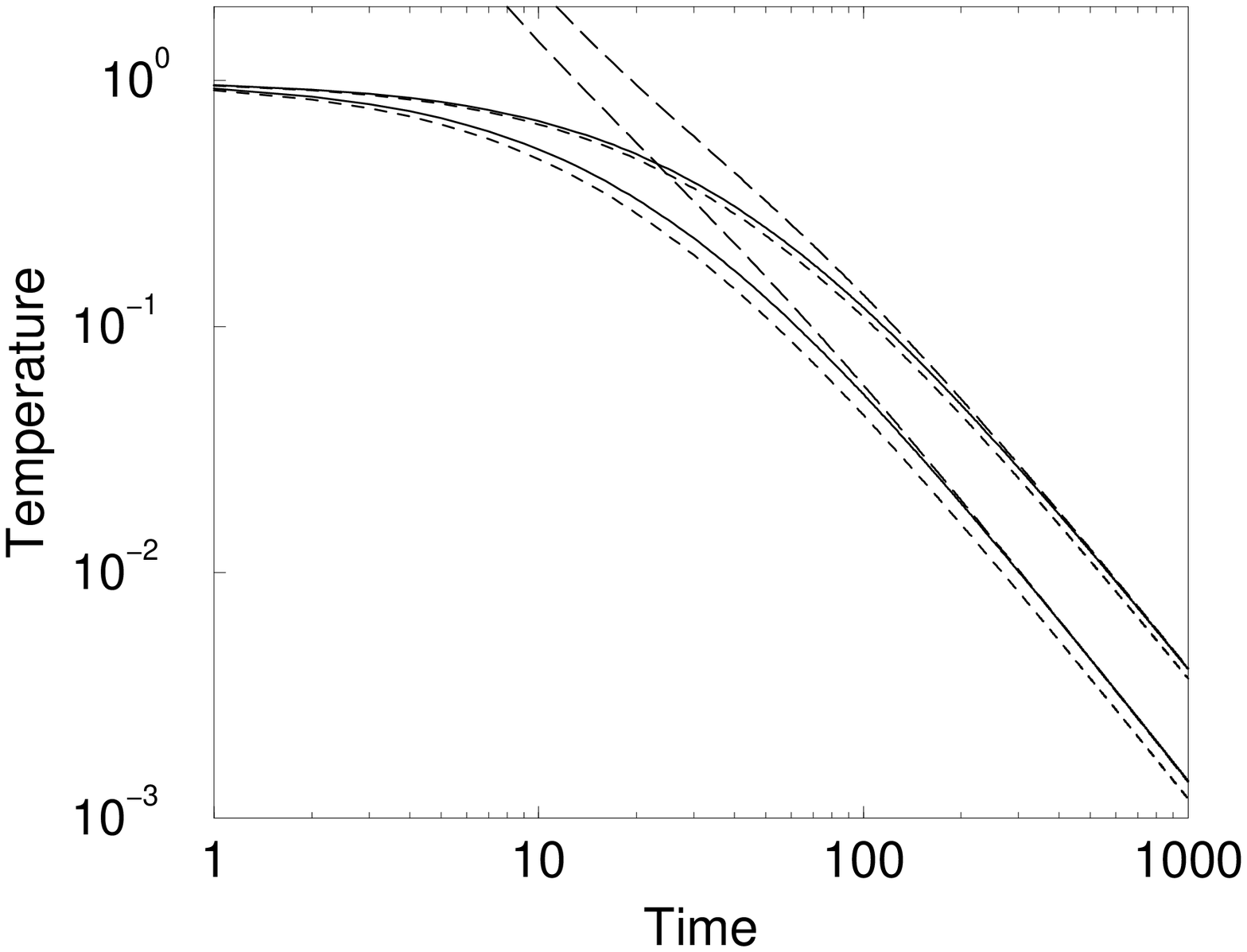,width=5.5cm}}  
  \caption{Time-evolution of the reduced temperature, $u(t)=T(t)/T_0$. 
The time is given in units of mean collisional time 
$\tau_c(0)$. Solid line: numerical solution, short-dashed:
$u_0(t)=(1+t/\tau_0)^{-5/3}$ (zero-order theory), 
long-dashed: $u(t)=u_0(t)+\delta \, u_1(t)$ (first-order theory). 
(Left): for $\delta=0.05, 0.1$ (top to bottom); (Right): $\delta=0.15, 0.25$ (top to bottom); 
(Middle): the same as (Left) but log-scale and larger ranges.}
  \label{fig:T}
\end{figure}

In Fig.~\ref{fig:a2} and Fig.~\ref{fig:T} we show the time dependence of the coefficient 
$a_2(t)$ of the Sonine polynomial expansion and of the temperature of the 
granular gas. The analytical findings are compared with the numerical 
solution of the system (\ref{genseteq1},\ref{genseteq2}). As one can see from the 
figures the analytical theory reproduces fairly well the 
numerical solution for the case of small $\delta$. 

As it follows from Fig.~\ref{fig:a2} (where the time is given  in collisional units), 
for small $\delta$ the following scenario of evolution of the velocity 
distribution takes place for a force-free granular gas. 
The initial Maxwellian distribution evolves to a non-Maxwellian distribution, with 
the discrepancy between these two characterized by the second coefficient of the 
Sonine polynomials expansion $a_2$. The deviation from the Maxwellian distribution 
(described by $a_2$) quickly grows, until it saturates after a few collisions per 
particle at a ``steady-state'' value. At this instant the deviation from the Maxwellian 
distribution is maximal, with the value $a_2 \approx - 0.4 \delta$ (Fig.~\ref{fig:a2}a). 
This refers to the first ``fast'' stage of the evolution, which takes place on a 
mean-collision time-scale $\sim \tau_c(0)$. After this maximal deviation is reached, 
the second ``slow'' stage of the evolution starts. At this stage $a_2$ decays to zero on 
the ``slow'' time scale $\tau_0 \sim \delta^{-1} \tau_c(0) \gg \tau_0(0)$, which 
corresponds to the time  scale of the temperature evolution (Fig.~\ref{fig:a2}b); the decay of the 
coefficient $a_2(t)$ in this regime occurs according to a power law $ \sim t^{-1/6}$ 
(Fig.~\ref{fig:a2}c). Asymptotically the 
Maxwellian distribution would be achieved, if the clustering process did not occur. 

Fig.~\ref{fig:T} illustrates the significance of the first-order correction $u_1(t)$ in the time-evolution 
of temperature. This becomes more important as the dissipation parameter $\delta$ 
grows (Figs.~\ref{fig:T}a,b). At large times the results of the first-order theory 
(with $u_1(t)$ included) practically coincide with the numerical results, while 
zero-order theory (without $u_1(t)$) demonstrates noticeable deviations (Fig.~\ref{fig:T}c). 

\begin{figure}[htbp]
\centerline{\psfig{figure=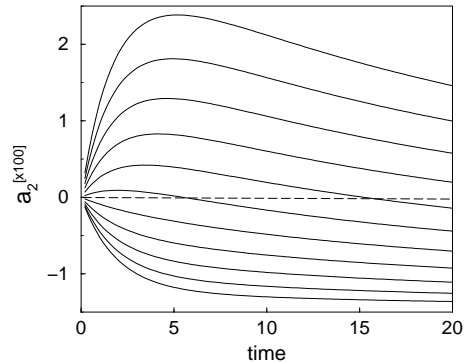,width=6cm}}  
  \caption{Time dependence of the second coefficient of the Sonine polynomial expansion $a_2(t) \times 100 $.
Time is given in units of mean collisional time $\tau_c(0)$. $\delta=0.1, 0.11, 0.12, \ldots, 0.20$ (bottom to top).}
  \label{fig:a2time}
\end{figure}

For larger values of $\delta$ the linear theory breaks down. Unfortunately, the 
equations obtained for the second order approximation ${\cal O}(\delta^2)$ are 
too complicated to be treated analytically. Hence, we studied them only numerically
(see Fig.~\ref{fig:a2time}). As compared to the case of small $\delta$, an additional intermediate regime in 
the time-evolution of the velocity distribution is observed. The first ``fast'' stage 
of evolution takes place, as before, on the time scale of few collisions per particle,
where maximal deviation from the Maxwellian distribution is achieved (Fig.~\ref{fig:a2time}). For 
$\delta \geq 0.15$ these maximal values of $a_2$ are  positive. Then, on the second 
stage (intermediate regime), which continues $10-100$ collisions, $a_2$ changes its 
sign and reaches a maximal negative deviation. Finally, on the third, slow stage, 
$a_2(t)$ relaxes to zero on the slow time-scale $\sim \tau_0$, just as for 
small $\delta$. In Fig.~\ref{fig:a2time} we show the first stage of the 
time evolution of $a_2(t)$ for systems  with large $\delta$. At a certain value of the dissipative parameter $\delta$ the behavior changes qualitatively, i.e. the system then reveals another time scale as discussed above. 

\begin{figure}[htbp]
\centerline{\psfig{figure=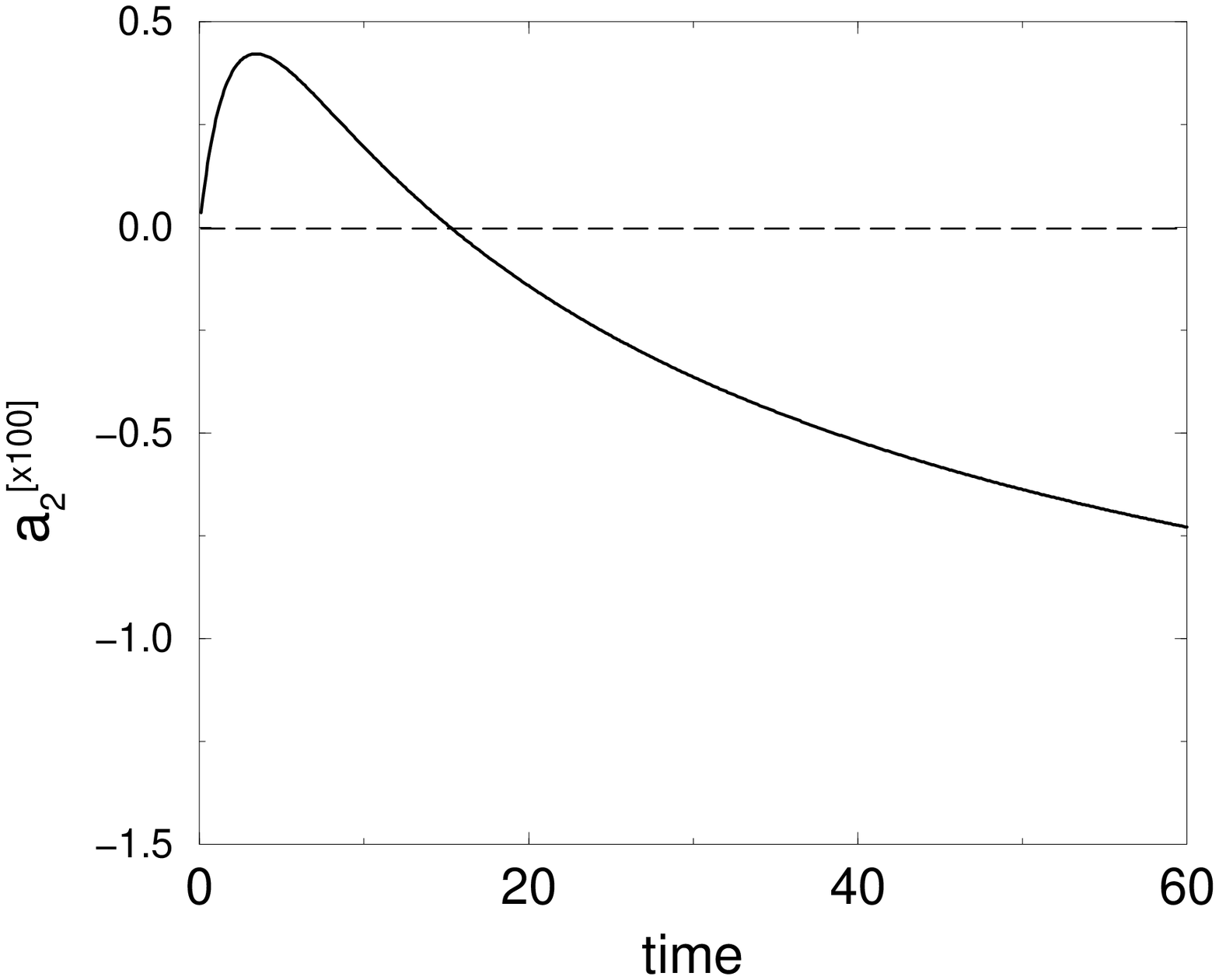,width=5cm}\psfig{figure=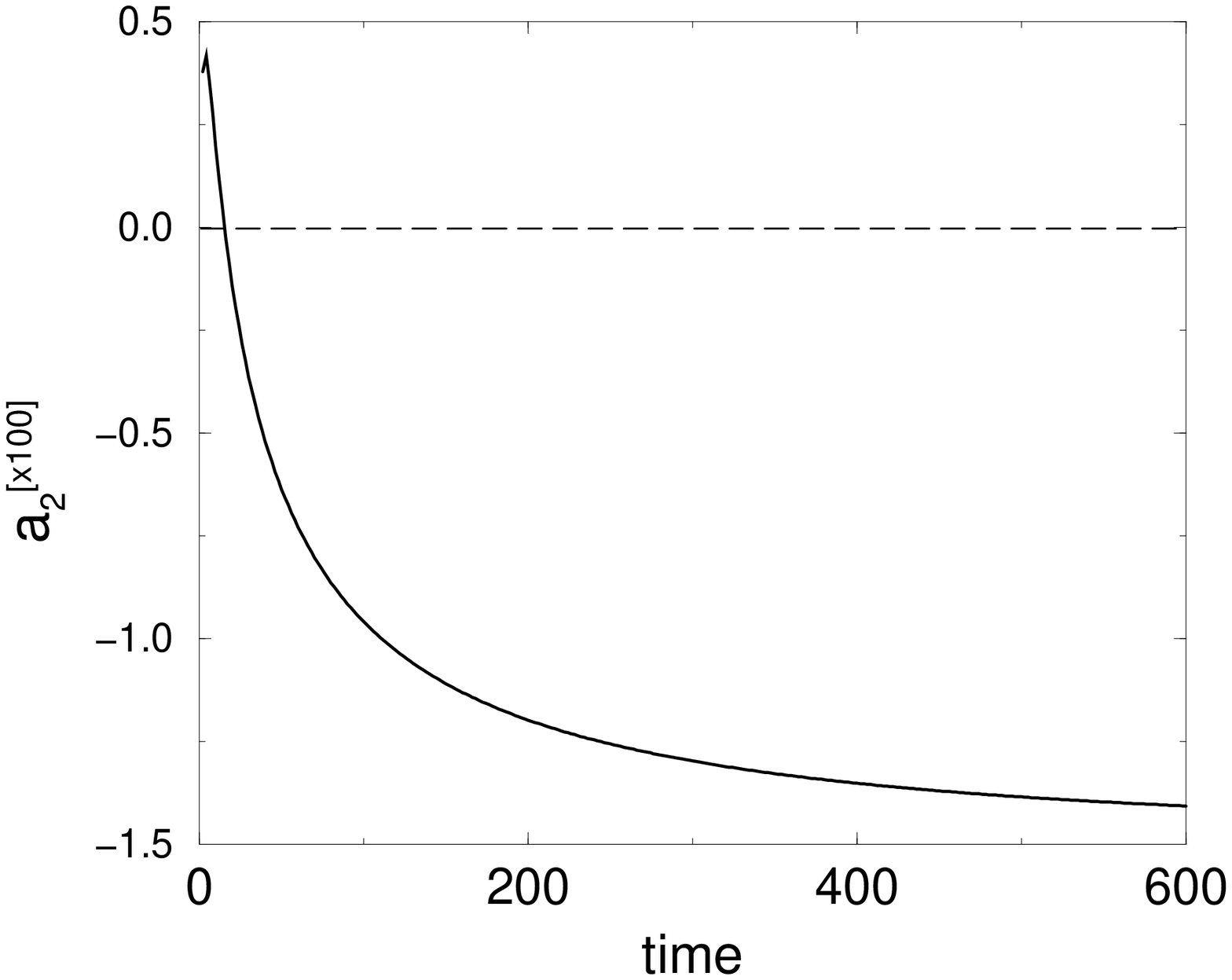,width=5cm}}
\centerline{\psfig{figure=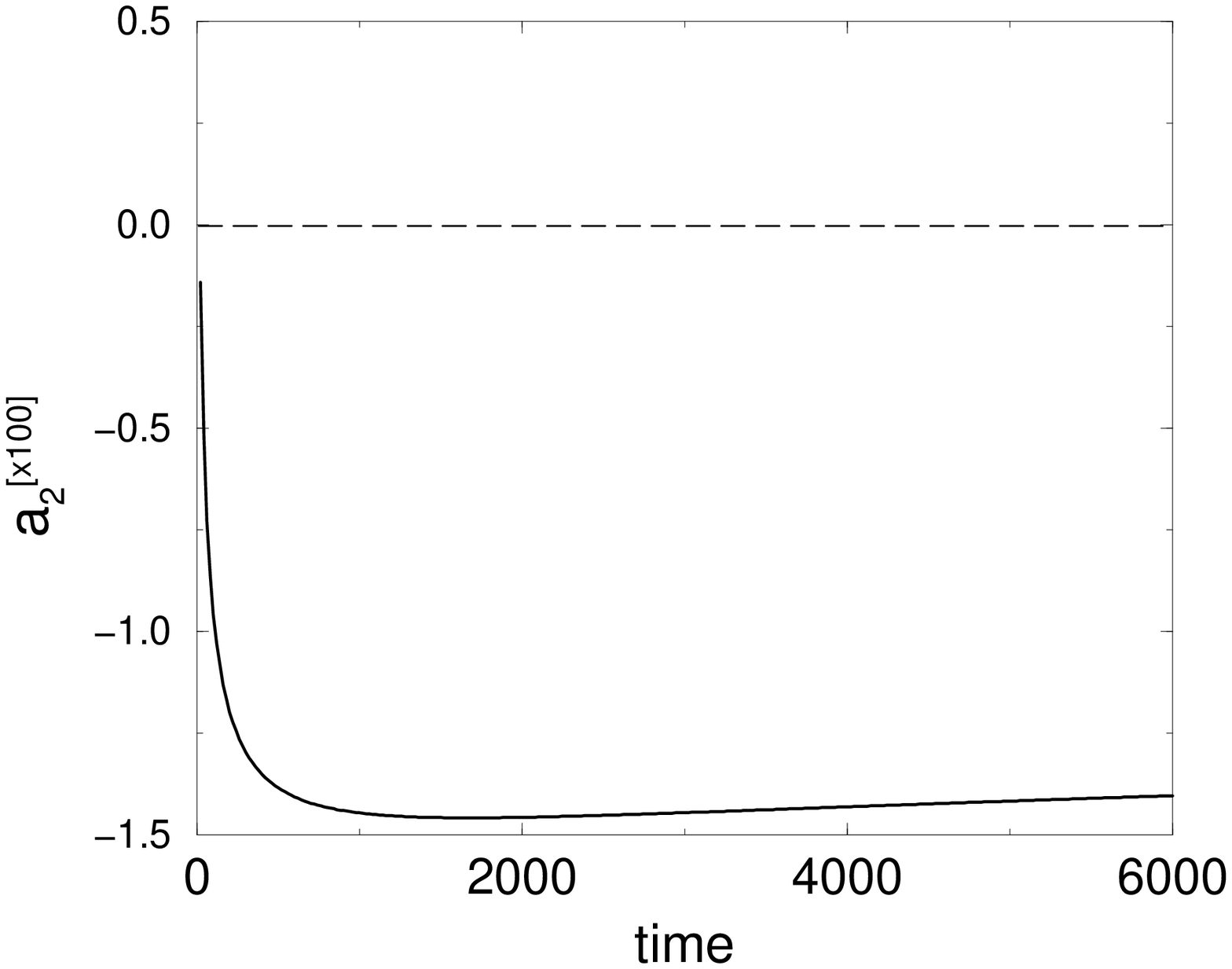,width=5cm}\psfig{figure=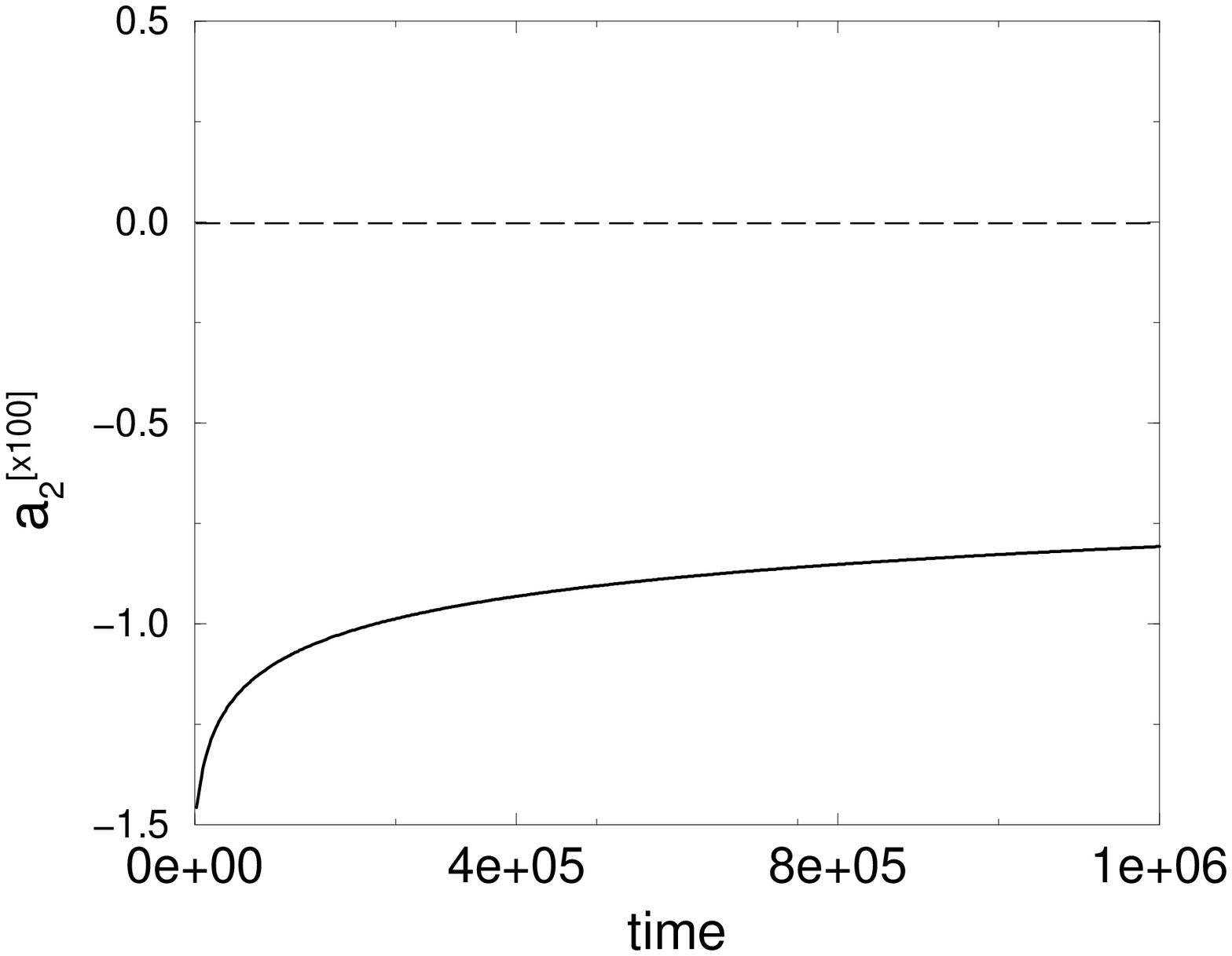,width=5cm}}
  \caption{The second Sonine coefficient $a_2$  for $\delta=0.16$ over time. The numerical solutions of Eqs.~(\ref{genseteq1}) and (\ref{genseteq2}) show all stages of evolution discussed in the text.}
  \label{fig:a2evo}
\end{figure}

Figure~\ref{fig:a2evo} shows the numerical solution of Eqs.~(\ref{genseteq1}) and (\ref{genseteq2}) for the second Sonine coefficient $a_2(t)$ as a function of time. One can clearly distinguish the different stages of evolution of the velocity distribution function. A more detailed investigation of the evolution of the distribution function for larger dissipation is subject of present research~\cite{BPMaxw}.

The interesting property of the granular gases in the regime of homogeneous 
cooling is the overpopulation of the high-velocity tails in the velocity 
distribution~\cite{EsipovPoeschel:97}, which has been shown for granular gases consisting of particles which interact via a constant restitution coefficient,
$\epsilon ={\rm const}$. How does the velocity dependence of the 
restitution coefficient as it appears for viscoelastic spheres influence this effect?  We observe, that for the 
case of $\epsilon=\epsilon(v_{\rm imp})$ the functional form (i.e. the exponential 
overpopulation~\cite{EsipovPoeschel:97}) persists, but it decreases with time on the 
``slow'' time-scale $\sim \tau_0$. Namely we obtain for the velocity distribution 
for $c \gg 1$ \cite{BPMaxw}:
\begin{equation}
\label{veldiscgg1}
\tilde{f}(\vec{c}, t) \sim 
\exp \left[ - \frac{b}{ \delta}\,c\left(1+\frac{t}{\tau_0} \right)^{1/6} \right]\, .
\end{equation}
where 
$b=\sqrt{\pi/2}\left(16q_0/5\right)^{-1}=2.25978\ldots$, which holds for 
$t \gg \tau_c(0)$.  
Again we see that the distribution tends asymptotically to the Maxwellian distribution, 
since the overpopulation vanishes as $t \to \infty$. 

Using the temperature and the velocity distribution of a granular gas as were derived in this section, one can 
calculate the kinetic coefficients. In the next section we consider the simplest one --
the self-diffusion coefficient. 

\section{Self-diffusion in granular gases of viscoelastic particles}

In the simplest case diffusion of particles occurs when there are 
density gradients in the system. 
The diffusion coefficient $D$ relates the flux of particles $\vec{J} $ to the 
density gradient $\nabla n$ according to a linear relation, provided the gradients are 
not too large:
\begin{equation}
\label{Jdiff}
\vec{J}= -D \nabla n\,.
\end{equation}
The coefficient $D$ also describes the statistical average of the  migration of a single particle. For {\em equilibrium} 
3D-systems the mean-square displacement of a particle reads
\begin{equation}
\left\langle \left( \Delta r(t) \right)^2  \right\rangle_{\rm eq} 
= 6\,D\,t \,,
\label{Dgen}
\end{equation}
where $\left< \cdots \right>_{\rm eq}$ denotes the {\em equilibrium}
ensemble averaging. For {\em nonequilibrium} systems, such as granular gases, 
one should consider the time-dependent diffusion coefficient $D(t)$ and the corresponding 
generalization of Eq.~(\ref{Dgen}):
\begin{equation}
\left\langle \left( \Delta r(t) \right)^2  \right\rangle =
6\, \int^t D(t^{\prime}) dt^{\prime}\,,
\label{Dgengen}
\end{equation}
where $\langle \cdots \rangle$ denotes averaging over the nonequilibrium ensemble. 
If the migration of a particle occurs in a uniform system composed of particles of the 
same kind, this process is called ``self-diffusion''. Correspondingly, the kinetic 
coefficient $D$ is called self-diffusion coefficient. 

To find  the mean-square  displacement, one writes 
\begin{equation}
\left\langle \left( \Delta r(t) \right)^2  \right\rangle =
\left< \int_0^{t}\vec{v}(t^{\prime})dt^{\prime}
\int_0^t  \vec{v}(t^{\prime \prime}) dt^{\prime \prime}  \right> \, 
\label{delR}
\end{equation}
and encounters then with the velocity autocorrelation function

$$
K_v(t^{\prime}, t) 
\equiv \left< \vec{v}(t^{\prime}) \vec{v}(t^{\prime \prime}) \right> 
$$
which should be evaluated in order to obtain the mean-square displacement and 
the self-diffusion coefficient. 

To calculate $K_v(t^{\prime}, t)$ we use the approximation of uncorrelated 
successive binary collision, which is valid for moderately dense systems, and 
an approach based on the formalism of the pseudo-Liouville operator 
${\cal L}$~\cite{pseudo}. The pseudo-Liouville operator is defined as
\begin{equation}
i{\cal L}=\sum_j \vec{v}_j \cdot \frac{\partial}{\partial \vec{r}_j}
+\sum_{i<j}\, \hat{T}_{\mbox{\footnotesize\em ij}}\,.
\label{L}
\end{equation}
The first sum in (\ref{L}) refers to the free streaming of the
particles (the ideal part) while the second sum refers to the particle
interactions which are described by the binary collision
operators~\cite{Chandler}
\begin{equation}
\hat{T}_{\mbox{\footnotesize\em ij}}\!=\sigma^{2}\!\! \int\!\! d\vec{e}\, 
\Theta \left(- \vec{v}_{\mbox{\footnotesize\em ij}} \cdot \vec{e}\, \right)\!
|\vec{v}_{\mbox{\footnotesize\em ij}} \cdot \vec{e}\, | 
\delta\! \left( \vec{r}_{\mbox{\footnotesize\em ij}}- \sigma \vec{e} 
\right)\!\!
\left(\hat{b}_{\mbox{\footnotesize\em ij}}^{\vec{e}}-1 \right)  \,,
\label{Tij}
\end{equation}
where $\Theta(x)$ is the Heaviside function.  The operator
$\hat{b}_{\mbox{\footnotesize\em ij}}^{{\vec{e}}}$ is defined as
\begin{equation}
\hat{b}_{\mbox{\footnotesize\em ij}}^{\vec{e}} f \left (\vec{r}_{i},
  \vec{r}_{j}, \vec{v}_{i},\vec{v}_{j} \cdots \right)=f \left
  (\vec{r}_{i}, \vec{r}_{j},
  \vec{v}^{*}_{i},\vec{v}^{*}_{j} \cdots \right) \, , 
\end{equation}
where $f$ is some function of the dynamical variables and $\vec{v}^{*}_{i}$ and $\vec{v}^{*}_{j}$ are the postcollisional velocities from Eq.~(\ref{directcoll}). 
The pseudo-Liouville operator gives the time derivative of any
dynamical variable $B$ (e.g.~\cite{resibua}):
\begin{equation}
\frac{d}{dt} B\left( \left\{ \vec{r}_i, \vec{v}_i  \right\}, t \right)= 
i{\cal L}\, B\left( \left\{     \vec{r}_i, \vec{v}_i \right\}, t \right)\,.
\label{derA}
\end{equation}
Therefore, the  time evolution of $B$
reads ($t>t^{\prime}$)
\begin{equation}
B\left( \{ \vec{r}_i, \vec{v}_i  \}, t \right)=
e^{i{\cal L} (t-t^{\,\prime} )} 
B\left( \{     \vec{r}_i, \vec{v}_i \}, t^{\,\prime} \right)\,.
\label{evolA}
\end{equation}
With Eq.~(\ref{evolA}) 
the time-correlation function reads
\begin{equation}
\left< B(t^{\prime})B(t) \right >=
\int d\Gamma \rho(t^{\prime}) B(t^{\prime}) e^{i{\cal L} (t-t^{\prime})}
B(t^{\prime})\,,
\label{evolAA}
\end{equation}
where $\int d\Gamma$ denotes integration over all degrees of freedom
and $\rho(t^{\prime})$ depends on temperature $T$, density $n$, etc.,
which change on a time-scale $t \gg \tau_c$.

Now we assume that
\begin{enumerate}
\item[(i)] the coordinate part and the velocity part of the distribution 
   function $\rho(t)$ factorize, and 
\item [(ii)] the molecular chaos hypothesis is valid.
\end{enumerate}
This suggests the following form of the distribution function:
\begin{equation}
\rho(t)=\rho(\vec{r}_1, \ldots, \vec{r}_N)\cdot f(\vec{v}_1,t) \ldots f(\vec{v}_N,t)\,.
\label{molchaos}
\end{equation}
In accordance with the
molecular chaos assumption  the sequence of the
successive collisions occurs without correlations. If the variable $B$
does not depend on the positions of the particles, its
time-correlation function decays exponentially~\cite{Chandler1}:

\begin{equation}
\left\langle B(t^{\prime})B(t) \right \rangle =
\left< B^2 \right>_{t^{\prime}} e^{-\left.\left|t-t^\prime\right|\right/\tau_B(t^\prime)}
~~~~\left(t > t^{\prime}\right) \,.
\label{AAexp}
\end{equation}
where $\langle \cdots \rangle_{t^\prime}$ denotes the averaging with
the distribution function taken at time $t^{\prime}$.  The relaxation
time $\tau_B$ is inverse to the initial slope of the autocorrelation
function~\cite{Chandler1}, as it may be found from the time derivative
of $\left\langle B(t^{\prime} )B(t) \right \rangle$ taken at
$t=t^{\prime}$. Equations~(\ref{evolAA}) and (\ref{AAexp}) then yield
\begin{equation}
-\tau_B^{-1}(t^{\prime})=\int\!\!d\Gamma \rho(t^{\prime}) B {i{\cal L} }B /
\left\langle B^2 \right \rangle_{t^{\prime}}= 
\frac{\left\langle B i{\cal L}  B \right 
\rangle_{t^{\prime}}}{\left\langle B^2 \right \rangle_{t^{\prime}}}\,.
\label{ALA}
\end{equation}
The relaxation time $\tau_B^{-1}(t^{\prime})$ depends on time via
the distribution function $\rho(t^{\prime})$ and varies on the
time-scale $t \gg \tau_c$.

Let $B(t)$ be the velocity of some particle, say $\vec{v}_1(t)$.  Then
with $3T(t)=\left\langle v^2 \right \rangle_t$, 
Eqs.~(\ref{AAexp}) and (\ref{ALA}) (with Eqs.~(\ref{L}) and (\ref{Tij})) read~\cite{BrilliantovPoeschel:1998d}
\begin{equation}
\left\langle \vec{v}_1 (t^{\prime})\cdot  \vec{v}_1(t)\right\rangle =
3T(t^{\prime}) e^{-|t-t^{\prime}|/\tau_v(t^{\prime})}
\label{vvexp}
\end{equation} 
\begin{equation}
-\tau_v^{-1}(t^\prime)=
(N-1) \frac{\left< \vec{v}_1 \cdot \hat{T}_{12} 
\vec{v}_1\right>_{t^{\prime}}}{\left< \vec{v}_1  \cdot \vec{v}_1 \right>_{t^{\prime}}}\,.
\label{vTv}
\end{equation} 
To obtain Eq.~(\ref{vTv}) we take into account that ${\cal L}_0
\vec{v}_1=0$, $\hat{T}_{\mbox{\footnotesize\em ij}}\,\vec{v}_1=0$ (for $i\neq 1$) and the identity of the particles. 

Straightforward calculation yields for the case of a constant 
restitution coefficient:
\begin{equation}
\tau_v^{-1}(t)=\frac{\epsilon +1}{2}\frac83 n \sigma^2 g_2(\sigma) 
\sqrt{\pi T(t)} =\frac{\epsilon +1}{2}  \tau_E^{-1}(t)\,, 
\label{tEt}
\end{equation}
where $\tau_E(t)=\frac32\,\tau_c(t)$ is the Enskog relaxation time
\cite{resibua}. Note that according to Eq.~(\ref{tEt}),   
$\tau_v =\frac{2}{1+\epsilon} \tau_E > \tau_E$, i.e., 
the velocity correlation time for inelastic collisions exceeds that of
elastic collisions. This follows from partial suppression of the
backscattering of particles due to inelastic losses in their normal
relative motion, which, thus, leads to more stretched particle trajectories, 
as compared to the elastic case.

Similar (although somewhat more complicated) computations may be 
performed for the system of viscoelastic 
particles yielding
\begin{equation}
\tau_v^{-1}(t)= 
\tau_E^{-1}(t) \left[1+\frac{3}{16}a_2(t)-4q_0 \, \delta \,u^{1/10}(t) \right]\,,
\label{tauva}
\end{equation} 
where $q_0=0.173318$ has been already introduced and $a_2(t)$, $u(t)$ are the 
same as defined above.
To obtain Eq.~(\ref{tauva}) we neglect terms of the order 
of ${\cal O}(a_2^2)$, 
${\cal O}(\delta^2)$ and ${\cal O}(a_2\, \delta)$. 

Using the velocity correlation function one writes
\begin{equation}
\left\langle \left( \Delta r(t) \right)^2  \right\rangle 
=2 \int_0^t dt^\prime 3T (t^\prime) \int_{t^\prime}^t 
dt^{\prime\prime} e^{-|t^{\prime\prime}-t^\prime|/\tau_v(t^\prime)}\,.
\label{Difvel}
\end{equation}
On the short-time scale $t \sim \tau_c$, $T (t^\prime)$ and
$\tau_v(t^\prime)$ may be considered as constants. Integrating in
(\ref{Difvel}) over $t^{\prime\prime}$ and equating with
(\ref{Dgengen}) yields for $t \gg \tau_c \sim \tau_v$ the
time-dependent self-diffusion coefficient
\begin{equation}
D(t)= T(t) \tau_v(t)\,.
\label{Difviatau}
\end{equation}

Substituting the dependencies for $u(t)=T(t)/T_0$ and $a_2(t)$ as functions of time, which has been derived in 
the previous section, we obtain the time dependence of the coefficient of 
self-diffusion $D(t)$. For $t \gg \tau_0$ this may be given in an explicit form:
\begin{equation}
\frac{D(t)}{D_0} \simeq \left(\frac{t}{\tau_0} \right)^{-5/6} \!\!
+\delta \left(4q_0+q_1+\frac{21}{400}h \right) \left(\frac{t}{\tau_0} \right)^{-1}\,,
\label{Diffinal}
\end{equation} 
where the constants $q_0$, $q_1$ and $h$ are given above. Hence,
the prefactor in the term proportional to $\delta$ reads 
$\left(4q_0+q_1+\frac{21}{400}h \right)=2.24956$, and $D_0$ is the initial Enskog value 
of the self-diffusion coefficient
\begin{equation}
D_0^{-1} =  \frac83 \, \pi^{1/2} n g_2(\sigma) \sigma^2 T_0^{-1/2}\,.
\end{equation}  
Correspondingly, the mean-square displacement reads
asymptotically for $t \gg \tau_0 $:
\begin{equation}
\left\langle \left( \Delta r(t) \right)^2  \right\rangle  \sim 
t^{1/6}+ b \, \delta\, \log t + \ldots\,,
\label{dRasym}
\end{equation}
where $b$ is some constant.
This dependence holds true for times 
\begin{equation}
\tau_c(0)\,\delta^{-1} \ll t \ll \tau_c(0)\, \delta^{-11/5}\,.
\label{ineq}
\end{equation}
The
first inequality in Eq.~(\ref{ineq}) follows from the condition $\tau_0 \ll t$, while the
second one follows from the condition  $\tau_c(t) \ll
\tau_0$, which means that temperature changes are slow on the collisional time-scale. For the constant restitution coefficient one obtains
\begin{equation}
T(t)/T_0=\left[ 1+\gamma_0 t/\tau_c(0) \right]^{-2}\,,  
\end{equation}
where $\gamma_0 \equiv \left(1-\epsilon^2\right)/6$ \cite{GZ93,NEBO97}. Thus, using 
Eqs.~(\ref{tEt}) and  (\ref{Difviatau}) one obtains for the
mean-square displacement in this case
\begin{equation}
\label{drcons}
\left< \left( \Delta r(t) \right)^2
\right>  \sim \log t \, .
\end{equation}
As it follows from Eqs.~(\ref{dRasym}) and (\ref{drcons}) the impact-velocity 
dependent restitution coefficient, Eq.~(\ref{epsC1C2}), leads to a significant 
change of the long-time behavior of the mean-square displacement of 
particles in the laboratory-time. Compared to its logarithmically 
weak dependence for the constant restitution coefficient, the 
impact-velocity dependence of the restitution coefficient
gives rise to a considerably faster increase of this quantity with time, 
according to a power law. 

One can also compare the dynamics of the system in its 
inherent-time scale. First we consider the average cumulative number of 
collisions per particle ${\cal N}(t)$ as an inherent measure for time 
(e.g.~\cite{McNamara96,EBEuro}). 
 It may be found by integrating 
$d{\cal N}=\tau_c(t)^{-1}dt$ \cite{NEBO97}. For a constant restitution 
coefficient $\epsilon$ one obtains ${\cal N}(t) \sim \log t$, while for the impact-velocity 
dependent $\epsilon\left( v_{\rm imp} \right)$ one has ${\cal N}(t) \sim t^{1/6}$. Therefore, the temperature and the mean-square
displacement behave in these cases as\\

\begin{center}
\renewcommand{\arraystretch}{1.4}
\setlength\tabcolsep{5pt}
\begin{tabular}{l|l}
\hline
$\epsilon={\rm const}$ & $\epsilon=\epsilon\left( v_{\rm imp} \right)$\\ 
\hline\noalign{\smallskip}                   
$T({\cal N}) \sim e^{-2(1-\epsilon^2){\cal N}}$ & $T({\cal N}) \sim {\cal N}^{-10} $ \\ 
\noalign{\smallskip}
\hline
\noalign{\smallskip}                
 $\left< \left( \Delta r({\cal N}) \right)^2\right>  \sim {\cal N}$ & $\left< \left( \Delta r({\cal N}) \right)^2\right>  \sim {\cal N}$ \\
\noalign{\smallskip}                
\hline
\end{tabular}
\end{center}
\vspace{0.1cm}

 If the number of collisions per particle ${\cal N}(t)$ would be the relevant
quantity specifying the stage of the granular gas evolution, one would
 conjecture that the dynamical behavior of a granular gas with a
constant $\epsilon$ and velocity-dependent $\epsilon$ are identical,
provided an ${\cal N}$-based time-scale is used.  Whereas in equilibrium systems the number of collisions is certainly an appropriate measure of time, in nonequilibrium systems this value has to be treated with more care. As a trivial example may serve a particle bouncing back and forth between two walls, each time it hits a wall it loses part of its energy: If one describes this system using a ${\cal N}$-based time, one would come to the conclusion that the system conserves its energy, which is certainly not the proper description of the system. According to our
understanding, therefore, the number of collision is not an appropriate time scale to describe physical reality. Sometimes, it may be even misleading.

Indeed, as it was shown in Ref.~\cite{EBEuro}, the value of
${\cal N}_c$, corresponding to a crossover from the linear regime of
evolution (which refers to the homogeneous cooling state) to the nonlinear
regime (when clustering starts) may differ by orders of magnitude,
depending on the restitution coefficient and on the density of the granular
gas.  Therefore, to analyze the behavior of a granular gas, one can try
an alternative inherent time-scale, ${\cal T}^{-1} \equiv T(t)/T_0$
which is based on the gas temperature. Given two
systems of granular particles at the same density and the same initial
temperature $T_0$, consisting of particles colliding with constant and
velocity-dependent restitution coefficient, respectively, the time
${\cal T}$ allows to compare directly their evolution. A strong
argument to use a temperature-based time has been given by Goldhirsch
and Zanetti~\cite{GZ93} who have shown that as the temperature decays, the 
evolution of the system changes from a linear regime to a nonlinear one. 
Recent numerical results of Ref.~\cite{EBEuro}
also support our assumption: It was shown that while ${\cal N}_c$
differs by more than a factor of three for two different systems, 
the values of ${\cal T}_c$, (defined, as ${\cal T}_c=T({\cal N}_c)/T_0$) are very close~\cite{EBEuro}.  These arguments show that one could consider ${\cal T}$ as a 
relevant time-scale to analyze the granular gas evolution.

With the temperature decay $T({\cal N})/T_0 \sim e^{-2 \gamma_0{\cal N}}$ for a constant 
restitution coefficient and $T({\cal N})/T_0 \sim {\cal N}^{-10}$ for the 
impact-velocity dependent one, we obtain the following dependencies:\\
\begin{center}
\renewcommand{\arraystretch}{1.4}
\setlength\tabcolsep{5pt}
\begin{tabular}{l|l}
\hline\noalign{\smallskip}
 $\epsilon={\rm const}$ & $\epsilon=\epsilon\left( v_{\rm imp} \right)$\\ 
\noalign{\smallskip}
\hline
\noalign{\smallskip}                   
 $T \sim \frac{1}{ {\cal T} }$ & $T \sim \frac{1}{ {\cal T} } $ \\ 
\noalign{\smallskip}
 $\left< \left( \Delta r({\cal T}) \right)^2\right>  \sim \log {\cal T}$ & $\left< \left( \Delta r({\cal T}) \right)^2 \right>  \sim {\cal T}^{1/10}$ \\
\noalign{\smallskip}
\hline
\end{tabular}
\end{center}
\vspace{0.1cm}

This shows that in the temperature-based time-scale, in which the  cooling 
of both systems is synchronized, the mean-square displacement grows logarithmically 
slow for the case of constant restitution coefficient and much faster, as a power 
law, for the system of viscoelastic particles with  $\epsilon = \epsilon(v_{\rm imp})$.
Thus, we conclude that clustering may be retarded for the latter system. 

\section{Conclusion}
In conclusion, we considered kinetic properties of granular gases composed of 
viscoelastic particles, which implies the 
impact-velocity dependence of the restitution coefficient. We found that 
such dependence gives rise to some new effects in granular gas dynamics:
(i) complicated, non-monotonous time-dependence of the coefficient $a_2$ of 
the Sonine polynomial expansion, which describes the deviation of the 
velocity distribution from the Maxwellian and (ii) enhanced spreading of 
particles, which depends  on time as a power law, compared to a logarithmically 
weak dependence  for the systems with a constant $\epsilon$. 

The Table below compares the properties of granular gases consisting of particles interacting via a constant coefficient of restitution $\epsilon ={\rm const}$ and consisting of viscoelastic particles where the collisions are described using an impact velocity dependent restitution coefficient $\epsilon = \epsilon(v_{\rm imp})$:\\
\begin{center}
\renewcommand{\arraystretch}{1.4}
\setlength\tabcolsep{5pt}
\begin{tabular}{l|l}
\hline\noalign{\smallskip}
 $\epsilon={\rm const}$ & $\epsilon=\epsilon\left( v_{\rm imp} \right)$\\ 
\noalign{\smallskip}
\hline
\noalign{\smallskip}                   
$\epsilon $  is a model parameter & $\epsilon=1-C_1A \kappa^{2/5}v_{\rm imp}^{1/5}+\cdots $ \\ 
 &$C_1=1.15396$, $C_2=\frac35 C_1^2$, $\ldots$ \\
 &$\kappa=\kappa(Y, \nu, m, R)$ \\
 &$A=A(\eta_1, \eta_2, Y, \nu)$ \\
 &all quantities are defined via parameters \\[-0.15cm]
 &of the particle material $Y$, $\nu$, $\eta_{1/2}$ \\[-0.15cm]
 &and their mass and radius. \\
\noalign{\smallskip}
\hline
\noalign{\smallskip}
\multicolumn{2}{c}{\bf Small parameter} \\
 $1-\epsilon^2$ -- does not depend & $\delta =A\kappa^{2/5}T_0^{1/10}$ -- depends \\[-0.15cm]
on the state of the system & on the initial temperature $T_0$.  \\
\noalign{\smallskip}
\hline
\noalign{\smallskip}
\multicolumn{2}{c}{\bf Temperature}\\
 $T=T_0 \left( 1+t/\tau_0^{\prime}\right)^{-2}$ & $T=T_0\left( 1+t/\tau_0\right)^{-5/3}$ \\
\noalign{\smallskip}
\hline
\noalign{\smallskip}
\multicolumn{2}{c}{\bf Velocity distribution}\\
 $f(\vec{v}, t)=\frac{n}{v_0^3(t)} \tilde{f}(\vec{c})$
& $f(\vec{v}, t)=\frac{n}{v_0^3(t)}\tilde{f}(\vec{c}, t) $ \\ 
$\tilde{f}(\vec{c},t)=\phi(c) \left\{1 + \sum_{p=1}^{\infty} a_p S_p(c^2) \right\}$ &
$\tilde{f}(\vec{c})=\phi(c) \left\{1 + \sum_{p=1}^{\infty} a_p(t) S_p(c^2) \right\}$ \\
 $a_2 ={\rm const.} $ & $a_2=a_2(t)$ -- is a (complicated) \\[-0.15cm]
& function of time. \\
\noalign{\smallskip}
\hline
\noalign{\smallskip}
\multicolumn{2}{c}{\bf Self-diffusion}\\
 $\left< \left( \Delta r(t) \right)^2\right>  \sim \log t $ & $\left< \left( \Delta r(t) \right)^2 \right>  \sim t^{1/6} $ \\
\noalign{\smallskip}
\hline
  \end{tabular}
\end{center}
\vspace{0.1cm}

\section*{Acknowledgements}

We thank M. H. Ernst and I. Goldhirsch for valuable discussions.


\begin{thebibliography}{99}
\addcontentsline{toc}{section}{References}
\bibitem{BSHPnumer}
Y. Du, H. Li, and L. P. Kadanoff, Phys. Rev. Lett. {\bf 74}, 1268 (1995);
T. Zhou and L. P. Kadanoff, Phys. Rev. E {\bf 54}, 623 (1996);
A. Goldshtein, M. Shapiro, and C. Gutfinger, J. Fluid Mech. {\bf 316}, 29 (1996);
A. Goldshtein, V. N. Poturaev, and I. A. Shulyak, Izvestiya Akademii Nauk SSSR, Mechanika Zhidkosti i Gaza {\bf 2}, 166 (1990);
J. T. Jenkins and M. W. Richman, Arch. Part. Mech. Materials {\bf 87}, 355 (1985);
V. Buchholtz and T. P\"oschel, Granular Matter {\bf 1}, 33 (1998);
E. L. Grossman, T. Zhou, and E. Ben-Naim, Phys. Rev. E {\bf 55}, 4200 (1997);
F. Spahn, U. Schwarz, and J. Kurths, Phys. Rev. Lett. {\bf 78}, 1596 (1997);
S. Luding and H. J. Herrmann, Chaos {\bf 9}, 673 (1999);
S. Luding and S. McNamara, Gran. Matter {\bf 1}, 113 (1998);
T. P. C. van Noije, M. H. Ernst, and R. Brito, Phys. Rev. E {\bf 57}, R4891 (1998);
J. A. C. Orza, R. Brito, T. P. C. van Noije, and M. H. Ernst,  
Int. J. Mod. Phys. C {\bf 8}, 953 (1997);
N. Sela and I. Goldhirsch, J. Fluid. Mech., {\bf 361}, 41 (1998);
N. Sela, I. Goldhirsch, and H. Noskowicz, Phys. Fluids {\bf 8}, 2337 (1996);
I. Goldhirsch, M.-L. Tan, and G. Zanetti, J. Sci. Comp. {\bf 8}, 1 (1993);
T. Aspelmeier, G. Giese, and A. Zippelius, Phys. Rev. E {\bf 57}, 857 (1997);
J. J. Brey and D. Cubero, Phys. Rev. E {\bf 57}, 2019 (1998);
J. J. Brey, J. W. Dufty, C. S. Kim, and A. Santos, Phys. Rev. E {\bf 58}, 4648 (1998);
J. J. Brey, M. J. Ruiz-Montero, and F. Moreno, Phys. Rev. E {\bf 55}, 2846 (1997);
J. J. Brey, D. Cubero, and M. J. Ruiz-Montero, Phys. Rev. E {\bf 59}, 1256 (1999);
V. Kumaran,
Phys.Rev.E, {\bf 59}, 4188, (1999);
  V. Garzo and J. W. Dufty, Phys. Rev. E, {\bf 59}, 5895, (1999).
\bibitem{McNamara96}
  S. McNamara and W. R. Young, Phys. Rev. E {\bf 53}, 5089  (1996).   

\bibitem{GZ93} I.~Goldhirsch and G.~Zanetti, {\em Phys.~Rev.~Lett.}, {\bf 70}, 1619 (1993).

\bibitem{NoijeErnst:97} 
  T. P. C. van Noije and M. H. Ernst, 
  Granular Matter, {\bf 1}, 57 (1998).
\bibitem{EsipovPoeschel:97} 
  S. E. Esipov and T. P\"{o}schel, J. Stat.
  Phys., {\bf 86}, 1385 (1997).
\bibitem{NoijeErnstRing} 
  T. P. C. van Noije, M. H. Ernst and R.Brito, 
  Physica A {\bf 251}, 266 (1998).
\bibitem{GoldshteinShapiro95}
  A. Goldshtein and M. Shapiro, 
  J. Fluid. Mech., {\bf 282}, 75 (1995). 
\bibitem{EBEuro}R. Brito and M. H. Ernst,  
  Europhys. Lett. {\bf 43}, 497 (1998).
\bibitem{NEBO97}T. P. C. van Noije and M. H.
  Ernst, R. Brito and J. A. G. Orza, Phys. Rev. Lett. {\bf 79}, 411
  (1997).  
\bibitem{BrilliantovPoeschel:1998d} 
  N. V. Brilliantov and T.
  P\"oschel, Phys. Rev. E {\bf 61}, 1716 (2000).
\bibitem{HuthmannZippelius:97} 
    M. Huthmann and A. Zippelius, { Phys.
    Rev. E.} {\bf 56}, R6275 (1997); S. Luding, M. Huthmann, 
    S. McNamara  and A. Zippelius, { Phys.
    Rev. E.} {\bf 58}, 3416 (1998). 

\bibitem{ThorntonHere} C. Thornton, {\em Contact mechanics and coefficients of restitution}, (in this volume, page 184.



\bibitem{rospap}
  R. Ram\'{\i}rez, T. P\"oschel, N. V. Brilliantov,
  and T. Schwager, Phys. Rev. E {\bf 60}, 4465 (1999).

\bibitem{Taguchi93} Y. Taguchi, Europhys. Lett. {\bf 24}, 203 (1993).

\bibitem{Luding98} S. Luding, {\em Collisions and Contacts between two particles}, in: H. ~J. Herrmann, J. -P. Hovi, and S. Luding (eds.), {\em Physics of dry granular media - NATO ASI Series E350}, Kluwer (Dordrecht, 1998), p. 285. 

\bibitem{BSHP1}
  N. V. Brilliantov, F. Spahn, J.-M. Hertzsch, and T.
  P\"oschel, Phys. Rev. E {\bf 53}, 5382 (1996).
\bibitem{BSHP2}
  J.-M. Hertzsch,  F. Spahn, and N. V. Brilliantov, J. Phys. II (France), 
  {\bf 5}, 1725 (1995).
\bibitem{remark2}
  Derivation of the dissipative force given in~\cite{BSHP1,BSHP2} for colliding 
  spheres may be straightforwardly generalized to obtain the relation 
  (\ref{eldiss}) (or (A17) in~\cite{BSHP1,BSHP2}) for colliding bodies of any 
  shape, provided that displacement field in the bulk of the material 
  of bodies in contact is a one-valued function of the compression 
  (see also~\cite{BrilPoshprep}).
\bibitem{BrilPoshprep}
  N. V. Brilliantov and T. P\"oschel, in preparation. 
  
\bibitem{LandauLifshits}
    L. D. Landau and E. M. Lifschitz, {\it Theory of Elasticity},
    Oxford University Press (Oxford, 1965). 
\bibitem{Hertz.Brill}
  H. Hertz, J. f. reine u. angewandte Math. {\bf 92}, 156 (1882).

\bibitem{Kuwabara} G. Kuwabara and K. Kono, Jpn. J. Appl. Phys. {\bf 26}, 1230 (1987).

\bibitem{TomThor}
  T. Schwager and T. P\"oschel, Phys. Rev. E
  {\bf 57}, 650 (1998).

\bibitem{resibua} P. Resibois and M. de Leener, {\em Classical Kinetic
    Theory of Fluids} (Wiley, New York, 1977).
\bibitem{CarnahanStarling}
  N. F. Carnahan, and K. E. Starling, J. Chem. Phys., {\bf 51}, 635 (1969). 
  
\bibitem{BPMaxw}
 N. V. Brilliantov and T. P\"oschel, Phys. Rev. E, {\bf 61}, 5573 (2000)

\bibitem{pseudo} 
  The term ``pseudo'' was initially referred to the
  dynamics of systems with singular hard-core potential
  \cite{Ernst:69,resibua}.
\bibitem{Chandler} For application to ``ordinary'' fluids, see
  \cite{Chandler1} and to granular systems
  \cite{NoijeErnstRing,HuthmannZippelius:97}.  A rigorous
  definition of ${\cal L}$ includes a pre-factor, preventing
  successive collisions of the same pair of particles
  \cite{Ernst:69,resibua} which, however, does not affect the present
  analysis.

\bibitem{Chandler1} 
  D. Chandler, {J. Chem. Phys.} {\bf 60}, 3500,
  3508 (1974); B. J. Berne, {\em ibid} {\bf 66}, 2821, (1977); 
  N. V. Brilliantov and O. P. Revokatov, 
  Chem. Phys. Lett. {\bf 104}, 444 (1984).

\bibitem{Ernst:69} 
  M. H. Ernst, J. R. Dorfman, W. R. Hoegy and J. M.
  J. van Leeuwen, Physica A {\bf 45}, 127 (1969).

\end{thebibliography}
\end{document}